\documentclass[12pt,preprint]{aastex}

\newcommand{\boldsymbol}[1]{\ensuremath{\mbox{\boldmath$#1$\unboldmath}}}
\newcommand{\underset}[2]{\ensuremath{{#2}{}{_{(#1)}}}}
\shorttitle{General relativistic model of light propagation: the dynamic case}
\shortauthors{de Felice et al.}

\begin{document}

\title{A general relativistic model of light propagation in the gravitational
field of the Solar System: the dynamical case}

\author{F. de Felice\altaffilmark{1}\altaffiltext{1}{INFN - Sezione di Padova}}
\affil{Department of Physics, University of Padova}
\affil{via Marzolo 8, 35131 Padova, Italy}
\email{fernando.defelice@pd.infn.it}

\author{A. Vecchiato}
\affil{INAF -- Turin Astronomical Observatory}
\affil{strada Osservatorio 20, 10125 Pino Torinese (TO), Italy}
\email{vecchiato@to.astro.it}

\author{M.~T. Crosta}
\affil{INAF -- Turin Astronomical Observatory}
\affil{strada Osservatorio 20, 10125 Pino Torinese (TO), Italy}
\email{crosta@to.astro.it}

\author{B. Bucciarelli}
\affil{INAF -- Turin Astronomical Observatory}
\affil{strada Osservatorio 20, 10125 Pino Torinese (TO), Italy}
\email{bucciarelli@to.astro.it}

\and

\author{M.~G. Lattanzi}
\affil{INAF -- Turin Astronomical Observatory}
\affil{strada Osservatorio 20, 10125 Pino Torinese (TO), Italy}
\email{lattanzi@to.astro.it}

\begin{abstract}
Modern astrometry is based on angular measurements at the micro-arcsecond level.
At this accuracy a fully general relativistic treatment of the data reduction is
required. This paper concludes a series of articles dedicated to the problem of 
relativistic light propagation, presenting the final microarcsecond version of a 
relativistic astrometric model which enable us to trace back the light path to 
its emitting source throughout the non-stationary gravity field of the moving 
bodies in the Solar System. The previous model is used as test-bed for
numerical comparisons to the present one. Here we also test different
versions of the computer code implementing the model at different levels of
complexity to start exploring the best trade-off between numerical efficiency
and the $\mu\textrm{as}$ accuracy needed to be reached.
\end{abstract}


\section{Introduction}
Modern space technology will soon provide stellar positioning with micro-arcsecond
accuracy ($\mu$as). At this level one has to take into account the general 
relativistic effects on light propagation arising from metric perturbations 
due not only to the bulk mass but also to the rotational and translational 
motion of the bodies of the Solar System, and to their multipole structure 
(see \citealp{2002PhRvD..65f4025K,2003AJ....125.1580K,2004CQGra..21.4463L}
and references therein). Our aim is to develop a Relativistic
Astrometric MODel (RAMOD) which enabled us to deduce, to the accuracy
of one $\mu$as, the astrometric parameters of a star in our Galaxy
from observations taken by modern space-born astrometric satellites like Gaia 
\citep{2005tdug.conf.....T}, which are fully consistent with the
precepts of General Relativity.

In this paper we present an astrometric model which contains an extension
to the dynamical case, {\em i.e.} with the inclusion of the $1/c^{3}$ terms,
of our previous model which was only accurate to $1/c^{2}$ 
\citep{2004ApJ...607..580D}. We term it as RAMOD3 since it was intended as 
the successor of two previous models (\citealt{1998AAp...332.1133D,2001AAp...373..336D}).
Following the same scheme, we shall refer to the model described here as RAMOD4.

The inclusion of terms of the order of $1/c^{3}$ corresponds
to an accuracy of $0.1~\mu$as, at least one order of magnitude better than the 
expected precision of the Gaia measurements. Here we
require that the Solar System is isolated and source of a weak gravitational
field. These two conditions imply that the velocities of the gravitational
sources within the Solar System are very small compared to the velocity
of light, typically of the order of $10\,$km s$^{-1}$. Under these
conditions we select a coordinate system $(x^{i},x^{0}\equiv ct)_{(i=1,\,2,\,3)}$
such that the background geometry has a post-Minkowskian form:
\begin{equation}
  g_{\alpha\beta}=\eta_{\alpha\beta}+h_{\alpha\beta}+O\left(h^{2}\right)
\label{eq:metric}
\end{equation}
where $\eta_{\alpha\beta}$ is the Minkowski metric and $h_{\alpha\beta}$
are small perturbations which describe effects generated by the bodies
of the Solar System. These perturbations are \textit{small} in the
sense that $|h_{\alpha\beta}|\ll1$, their spatial variations are
of the order of $|h_{\alpha\beta}|$ while their time variations are
of the order of $|h_{\alpha\beta}|/c$; here and in what follows Greek
indeces run from $0$ to $3$. Clearly metric form (\ref{eq:metric})
is preserved under coordinate gauge transformations of the order of
$h$. The $h_{\alpha\beta}$'s contain terms of the order of at least
$1/c^{2}$, hence we shall keep our approximation to first order in
$h$. To the order of $1/c^{3}$ the time dependence of the background
metric cannot be ignored therefore the time-like vector field 
$\boldsymbol{\eta}=\boldsymbol{\partial}_0$
tangent to the coordinate time axis, will not in general be a Killing 
field (namely an isometry for the space-time) unless one moves to far 
distances from the Solar System where the metric tends to be Minkowski's. 
The components of the vector field $\boldsymbol{\eta}$ are 
$\eta^{\mu}=\delta_{0}^{\mu},\,\eta_{\mu}=g_{0\mu}$ hence we easily 
deduce that the congruence $C_{\boldsymbol{\eta}}$, namely the family of curves 
having the vector field $\boldsymbol{\eta}$ as tangent field%
\footnote{These curves are also termed integral curves of the vector field 
$\boldsymbol{\eta}$.}, will have a non zero vorticity. From its definition 
\begin{equation}
  \omega_{\alpha\beta}=P(\eta)_{\alpha}{}^{\sigma}P(\eta)_{\beta}{}^{\rho}\nabla_{[\rho}\eta_{\sigma]},
\end{equation}
where $P(\eta)_{\lambda}^{\mu}=\delta_{\lambda}^{\mu}+\eta_{\lambda}\eta^{\mu}$
is the operator which projects orthogonally to $\boldsymbol{\eta}$,
$\nabla_{\rho}$ the covariant $\rho-$derivative relative to the
given metric and square brackets mean antisymmetrization, we have
that the only non vanishing components are, to the lowest order:
\begin{equation}
  \omega_{ij}=\partial_{[j}h_{i]0}+O\left(1/c^{4}\right),\quad i,j=1,\,2,\,3.
\label{eq:vort}
\end{equation}
We remind that $h_{0i}$ and $\partial_{j}h_{0i}$ are $\sim O(1/c^{3})$.
Condition (\ref{eq:vort}) implies that the surfaces $t=\mathrm{constant}$
are not orthogonal to the integral curves of $\boldsymbol{\eta}$
at least in and nearby the Solar System \citep{2004ApJ...607..580D}.
Nevertheless, because the time lines with tangent field $\boldsymbol{\eta}$
are asymptotically Killing and vorticity free then the slices $S(t):\; t=\mathrm{constant}$
allow for a non ambiguous $3+1$ splitting of spacetime with a coordinate
representation such that space-like coordinates are fixed within each
slice. We fix the origin of the space-like coordinates at the barycenter
of the Solar System and assume that the spatial coordinate axes point
to distant sources chosen in such a way to assure that the system
is kinematically non rotating. Adopting the IAU prescriptions \citep{2000IAU-res....B1.3},
this system is termed Barycentric Celestial Reference System (BCRS)
and will be our main system since all coordinate tensorial components
will be relative to it. The non stationarity of the spacetime makes
the construction of a relativistic astrometric model much less straightforward
than for the static case considered in \citet{2004ApJ...607..580D}.

In section~\ref{sec:light-trajectory} these complications are handled
to define the geometrical environment for light propagation and data
analysis. The spacetime metric is not diagonal since terms as $h_{0i}$
are different than zero; they are mainly generated by the velocity
of the metric sources relative to the given BCRS. Then the gravitational
potential at each point of the light trajectory depends on the sources
in the Solar System at the appropriate retarded position as specified
in section~\ref{sec:retarded-time}. This will have consequences
when fixing the observables and the boundary conditions for the differential
equations of the light rays in section~\ref{sec:boundary-conditions}.
Section~\ref{sec:Tests} shows how this model was numerically tested
and presents the results of these tests.

In what follows we shall use geometrized units such that $c=1=G$,
$G$ being the gravitational constant; with these units a mass $\mathcal{M}$,
expressed in kilograms, has the dimension of a length according to
the relation $M=G\mathcal{M}/c^{2}$; similarly the time coordinate
$x^{0}=ct$ will be simply written as $t$ and the spatial velocities
are in units of $c$. For sake of clarity, the velocity of light $c$
will appear explicitly only when we specify the order of magnitude
of terms under discussion.

\section{The light trajectory\label{sec:light-trajectory}}
\begin{figure}
  \plotone{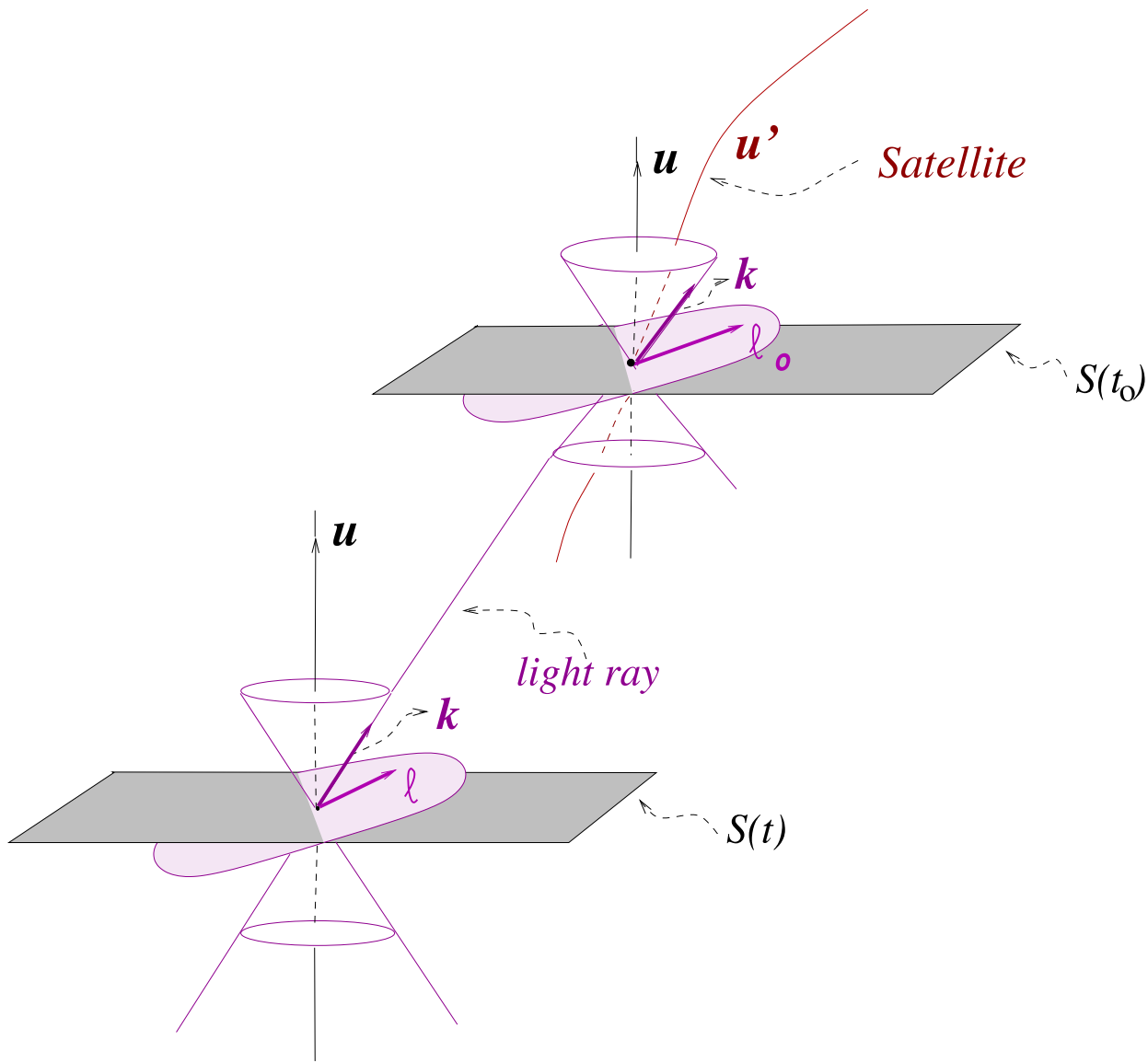}
  \caption{The light trajectory, identified by the four-vector $\boldsymbol{k}$,
    propagates in the space-time until it is intercepted by the Gaia-like 
    satellite at time $t_{0}$. At each point on its trajectory the light
    signal strikes the locally barycentric observer $\boldsymbol{u}$
    who identifies in its instantaneous rest-space (dotted area) the local
    line-of-sight $\boldsymbol{\ell}$. The surfaces $S(t):t=\mathrm{const}$
    do not in general coincide with the local rest-space of $\boldsymbol{u}$.}
\label{fig:k-split}
\end{figure}

We require that spacetime admits a family of hypersurfaces $S(t)$
with $t=\mathrm{constant}$ so that the spatial coordinates $\{ x^{i}\}$
are fixed on each of them. As said, these surfaces are constrained
by the condition of being asymptotically orthogonal to the \textit{time}
direction $\boldsymbol{\eta}$ which will also be asymptotically Killing.
In the nearby of, and of course within the Solar System, the spatial
coordinates $x^{i}$ are not constant along the normals to these hypersurfaces;
along them, in fact, they vary according to the shift law $\delta x^{i}\sim h_{0i}\delta t$.
To the order of $1/c^{3}$, all terms proportional to $h_{0i}$ are
in general not zero and cannot be made vanish in a gauge invariant
way. Let us term $\boldsymbol{u}$ a vector field parallel to $\boldsymbol{\eta}$
and tangent to the family of timelike curves $\hat{\gamma}$, say,
along which the spatial coordinates are constant; furthermore let
$\hat{\sigma}$ be the parameter on these curves which makes the vector
field $\boldsymbol{u}$ unitary, namely $u^{\alpha}u_{\alpha}=-1$.
Clearly along each integral curve of $\boldsymbol{u}$ carrying the
coordinates $x^{i}$, the parameter $\hat{\sigma}(x^{i},t)$ is function
of the coordinate time $t$ only. By definition, the vector field
$\boldsymbol{u}$ has components:
\begin{equation}
  \begin{array}{rcl}
    u^{\alpha} &=& \displaystyle{\frac{\mathrm{d}x^{\alpha}}{\mathrm{d}\hat{\sigma}}}
                   =\mathrm{e}^{\psi}\delta_{0}^{\alpha}\\
    u_{\alpha} &=& g_{0\alpha}\mathrm{e}^{\psi}
  \end{array}
\label{eq:hatu}
\end{equation}
where $\mathrm{e}^{\psi}=(\mathrm{d}t/\mathrm{d}\hat{\sigma})=(-g_{00})^{-1/2}$.
The integral curve of $\boldsymbol{u}$ through each spacetime point,
identifies a \textit{local barycentric observer} since, as stated,
this observer is at rest in the BCRS. Consider now a null geodesic
$\mit\Upsilon$ stemming from a distant star at the emission event
$P_{*}:(x_{*}^{i},\, t_{*})$ and ending at the observation event
$P_{0}:(x_{0}^{i},\, t_{0})$. Our aim is to find, from appropriate
observational data collected at $P_{0}$, the coordinate position
of the star and its proper motion. Let $\boldsymbol{k}$ be the vector
field tangent to $\mit\Upsilon$ and satisfying the following relations:
\begin{eqnarray}
  k_{\alpha}k^{\alpha} &=& 0\label{eq:nullcond}\\
  \frac{\mathrm{d}k^{\alpha}}{\mathrm{d}\lambda}+\Gamma^{\alpha}{}_{\mu\nu}k^{\mu}k^{\nu} 
    &=& 0\label{eq:geodes}
\end{eqnarray}
where (\ref{eq:geodes}) is the geodesic equation, $\lambda$ is
a real parameter on $\mit\Upsilon$ and $\Gamma^{\alpha}{}_{\mu\nu}$
are the connection coefficients of metric (\ref{eq:metric}) given
by:
\begin{equation}
  \Gamma^{\alpha}{}_{\mu\nu}=\frac{1}{2}\eta^{\alpha\rho}
    \left(\partial_{\mu}h_{\rho\nu}+\partial_{\nu}h_{\rho\mu}-
    \partial_{\rho}h_{\mu\nu}\right)+O\left(h^{2}\right).
\label{eq:connection}
\end{equation}
Condition (\ref{eq:vort}) holds for the family of the integral curves
of $\boldsymbol{u}$ as well, hence $C_{\boldsymbol{u}}$ does not
admit a global family of orthogonal hypersurfaces although such a
surface always exists locally, namely in a small neighborhood of each
point. This means that it is not possible to define a \textit{rest-space}
of the barycentric observer which covers the entire spacetime. The
slice $S(t_{0})$ for example is \textit{not} the rest-space of the
observer $\boldsymbol{u}$ at the observation point $P_{0}$. In fact,
neighboring events belonging to the local rest-space of $\boldsymbol{u}$
at $P_{0}$ and having spatial coordinates $x^{i}=x_{0}^{i}+\delta x^{i}$,
will have a coordinate time given by
\begin{equation}
  t=t_{0}+\mathrm{e}^{2\psi}h_{0j}\delta x^{j}.
\label{eq:deltati}
\end{equation}
The rest-space of the local barycentric observer $\boldsymbol{u}$
is locally identified by the operator $P(u)^{\alpha}{}_{\beta}=
\delta_{\beta}^{\alpha}+u^{\alpha}u_{\beta}$ which projects orthogonally 
to $\boldsymbol{u}$. We define the instantaneous \textit{line of sight} of 
$\boldsymbol{u}$ as the local spatial direction of light propagation; at 
each point of the photon path this direction is found projecting the tangent 
vector $\textbf{k}$ in the rest-space of $\boldsymbol{u}$ namely
\begin{equation}
  \ell^{\alpha}=P(u)^{\alpha}{}_{\beta}k^{\beta}=k^{\alpha}+u^{\alpha}(u_{\beta}k^{\beta}).
\label{eq:hatell}
\end{equation}
In this way we define a vector field $\boldsymbol{\ell}$ all along
the null curve $\mit\Upsilon$ (see figure~\ref{fig:k-split}); the
knowledge of $\ell^{\alpha}$ coupled with that of $u^{\alpha}$,
allows one to determine $k^{\alpha}$ and therefore to reconstruct
the null trajectory followed by the photon from the star to the satellite.
It is convenient to parameterize the null curve $\mit\Upsilon$ with
the quantity $\sigma(\lambda)\equiv\hat{\sigma}(x(\lambda),t)$ which
marks the proper time of the barycentric observer $\boldsymbol{u}$
which the light trajectory crosses at each $t$. %
\clearpage
\clearpage
If we set:
\begin{equation}
  \bar{\ell}^{\alpha}=-\frac{\ell^{\alpha}}{u^{\rho}k_{\rho}}
\label{eq:barell}
\end{equation}
we find from (\ref{eq:hatell}) that $\bar{\ell}_{\alpha}\bar{\ell}^{\alpha}=1$.
If we define similarly:
\begin{equation}
  \bar{k}^{\alpha}=-\frac{k^{\alpha}}{u^{\rho}k_{\rho}}\equiv\frac{\mathrm{d}x^{\alpha}}{\mathrm{d}\sigma}
\label{eq:bark}
\end{equation}
 then the equations which will solve the problem of finding the coordinate
position of the star at the emission, are fully determined if we set
along $\mit\Upsilon$
\begin{equation}
  \bar{\ell}^{\alpha}=\frac{\mathrm{d}x^{\alpha}}{\mathrm{d}\sigma}-u^{\alpha}.
\label{eq:xidot}
\end{equation}
From (\ref{eq:geodes}), (\ref{eq:connection}) and (\ref{eq:barell})
the differential equations of the local line of sight read:
{\arraycolsep=1pt
\begin{eqnarray}
\frac{\mathrm{d}\bar{\ell}^{\alpha}}{\mathrm{d}\sigma} &+&        
     \frac{1}{2}\left(\bar{\ell}^{0}\partial_{0}h_{00}\right)\delta_{0}^{\alpha}+
     \left(\bar{\ell}^{i}\partial_{i}h_{00}\right)\delta_{0}^{\alpha}+
     \frac{1}{2}\bar{\ell}^{\alpha}\bar{\ell}^{i}\partial_{i}h_{00}\nonumber \\
 &-& \left(\bar{\ell}^{\alpha}+\delta_{0}^{\alpha}\right)
     \left(\bar{\ell}^{i}\partial_{0}h_{0i}+\frac{1}{2}\bar{\ell}^{i}\bar{\ell}^{j}
     \partial_{0}h_{ij}\right)\nonumber \\
 &+& \eta^{\alpha\lambda}\left[\bar{\ell}^{\beta}\bar{\ell}^{\gamma}
     \left(\partial_{\beta}h_{\lambda\gamma}-
     \frac{1}{2}\partial_{\lambda}h_{\beta\gamma}\right)+\right.\nonumber \\
 & & \qquad\quad\bar{\ell}^{\beta}\left(\partial_{\beta}h_{\lambda0}+
     \partial_{0}h_{\lambda\beta}-\partial_{\lambda}h_{0\beta}\right)\bigg]\nonumber \\
 &+& \eta^{\alpha\lambda}\left(\partial_{0}h_{\lambda0}-
     \frac{1}{2}\partial_{\lambda}h_{00}\right)=0.
\label{eq:lighteq}
\end{eqnarray}
}

Let us recall an important property of the vector field $\bar{\boldsymbol{\ell}}$.
>From (\ref{eq:hatell}) and (\ref{eq:hatu}), we deduce that $\bar{\ell}_{0}=P(u)_{0\beta}k^{\beta}=0$;
this implies
\begin{equation}
  \bar{\ell}^{0}=h_{0i}\bar{\ell}^{i}+O(1/c^{4}),
\label{eq:ellzero}
\end{equation}
hence, to the order of $1/c^{3}$ equation (\ref{eq:lighteq}) becomes:
{\arraycolsep=1pt
\begin{eqnarray}
\frac{\mathrm{d}\bar{\ell}^{\alpha}}{\mathrm{d}\sigma} &+&
     \delta_{0}^{\alpha}\bar{\ell}^{i}\partial_{i}h_{00}+
     \frac{1}{2}\delta_{k}^{\alpha}\bar{\ell}^{k}\bar{\ell}^{i}\partial_{i}h_{00}\nonumber \\
 &-& \left(\bar{\ell}^{k}\delta_{k}^{\alpha}+\delta_{0}^{\alpha}\right)
      \frac{1}{2}\bar{\ell}^{i}\bar{\ell}^{j}\partial_{0}h_{ij}\nonumber \\
 &+& \eta^{\alpha\lambda}\left[\bar{\ell}^{i}\bar{\ell}^{j}
     \left(\partial_{i}h_{\lambda j}-\frac{1}{2}\partial_{\lambda}h_{ij}\right)+
\right.\nonumber \\
 & & \qquad\quad\bar{\ell}^{i}\left(\partial_{i}h_{\lambda0}+
     \partial_{0}h_{\lambda i}-\partial_{\lambda}h_{0i}\right)\bigg]\nonumber \\
 &-& \eta^{\alpha\lambda}\frac{1}{2}\partial_{\lambda}h_{00}=0
\label{eq:newlighteq}
\end{eqnarray}
}where $i,\, j,\, k\dots=1,\,2,\,3$. Property (\ref{eq:ellzero})
and the requirement that our analysis is to first order in $h$ allows
us to decouple the time component $\bar{\ell}^{0}$ from the spatial
components $\bar{\ell}^{i}$ in the set of differential equations
(\ref{eq:newlighteq}). We finally have to the $1/c^{3}$:
{\arraycolsep=1pt
\begin{eqnarray}
\frac{\mathrm{d}\bar{\ell}^{0}}{\mathrm{d}\sigma} &-& 
     \bar{\ell}^{i}\bar{\ell}^{j}\partial_{i}h_{0j}-
     \frac{1}{2}\partial_{0}h_{00}=0\label{eq:masterequation1}\\
\frac{\mathrm{d}\bar{\ell}^{k}}{\mathrm{d}\sigma} &-& 
     \frac{1}{2}\bar{\ell}^{k}\bar{\ell}^{i}\bar{\ell}^{j}
     \partial_{0}h_{ij}+\bar{\ell}^{i}\bar{\ell}^{j}
     \left(\partial_{i}h_{kj}-\frac{1}{2}\partial_{k}h_{ij}\right)\nonumber \\
 &+& \frac{1}{2}\bar{\ell}^{k}\bar{\ell}^{i}\partial_{i}h_{00}+
     \bar{\ell}^{i}\left(\partial_{i}h_{k0}+\partial_{0}h_{ki}-
     \partial_{k}h_{0i}\right)\nonumber \\
 &-& \frac{1}{2}\partial_{k}h_{00}=0.
\label{eq:masterequation2}
\end{eqnarray}
}
The boundary conditions needed to solve equations (\ref{eq:xidot}),
(\ref{eq:masterequation1}) and (\ref{eq:masterequation2}) are the
coordinate positions $x_{0}^{i}$ of the satellite with respect to
BCRS at the observation time $t_{0}$ and the local line of sight
direction $\bar{\ell}_{(0)}^{\alpha}$. While $x_{0}^{i}$ are supposed
to be known, the components $\bar{\ell}_{(0)}$ need to be expressed
in terms of specific observables which depend from the experimental
set-up.

\section{The retarded time corrections\label{sec:retarded-time}}
Integration of equations (\ref{eq:masterequation1}) and (\ref{eq:masterequation2}) 
obviously requires that one calculates the metric coefficients $h_{\alpha\beta}$ 
all along the integration path. Due to the linear regime, each perturbation term 
can be written as:
\begin{equation}
  h_{\alpha\beta}=\sum_{a}h_{\alpha\beta}^{(a)}
\label{eq:h}
\end{equation}
where the sum is extended to all bodies of the Solar System. To the 
selected order of $1/c^{3}$ and confining our attention to mass monopole 
terms of the gravitating sources, it is appropriate to consider a standard solution 
of Einstein's equations in terms of retarded tensor potential
(\citealp{1972gcpa.book.....W,1973grav.book.....M,1990recm.book.....D}), which can be 
specialized also as the Li\'enard-Wiechert ones \citep{1999PhRvD..60f124002K}. In 
the last case and in conventional units the metric coefficients read:
{\arraycolsep=2pt
\begin{equation}
  \left\{
  \begin{array}{rcl}
    h_{00}^{(a)}&=&\left(\displaystyle{\frac{2G\mathcal{M}^{(a)}}{c^{2}r_{_{\cal R}}^{(a)}}}\right)
                   +O\left(\displaystyle{\frac{1}{c^{4}}}\right) \\[15pt]
    h_{jk}^{(a)}&=&\left(\displaystyle{\frac{2G\mathcal{M}^{(a)}}{c^{2}r_{_{\cal R}}^{(a)}}}\right)\delta_{jk}
                   +O\left(\displaystyle{\frac{1}{c^{4}}}\right) \\[15pt]
    h_{0j}^{(a)}&=&\displaystyle{-\frac{2w_{j}^{(a)}}{c^{3}}}
                   +O\left(\displaystyle{\frac{1}{c^{4}}}\right),
  \end{array}
  \right.
\label{eq:hform}
\end{equation}
}where $w_j$ is a \textit{spatial} potential which describes the 
dynamical contribution to the background geometry by the relative 
motion of the gravitational sources and $r_{_{\cal R}}^{(a)}$, to be defined shortly, is the 
 distance from the points on the photon trajectory to the 
barycenter of the $a$-th gravity source \textit{at the appropriate 
retarded time}. The retarded position of the source is fixed by the intercept of its world-line 
with the past light-cone at any point of the photon trajectory.
 We shall now calculate this distance warning the reader 
that we shall use  geometrized units ($c=1=G$) again.

\begin{figure}
  \plotone{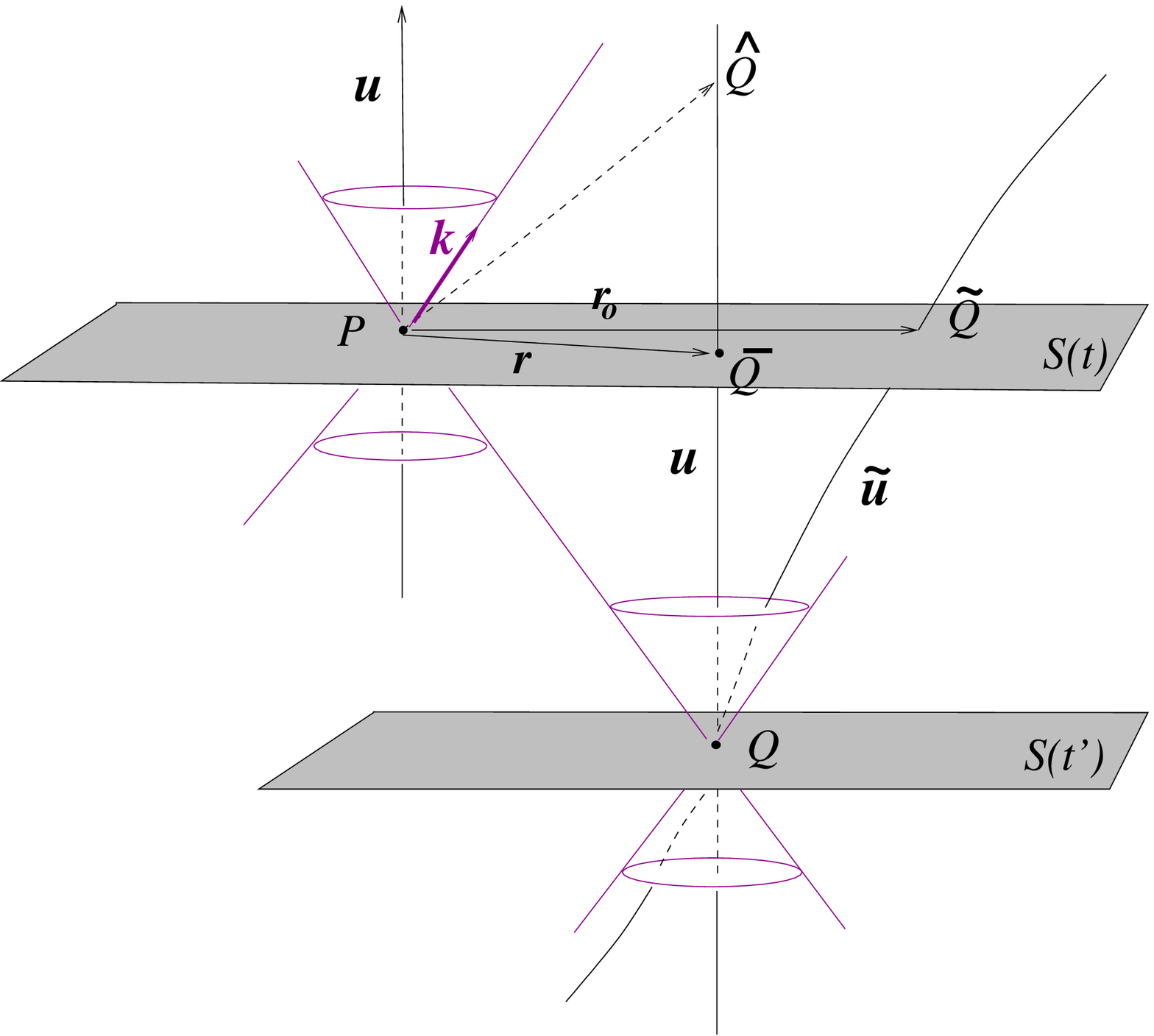}
  \caption{\label{fig:ret-time}The metric coefficients in $P$ at time $t$
    are determined by the gravity source located in $Q$ at the retarded
    time $t'$. At time $t$ the gravity source will be in $\tilde{Q}$
    but the distance $r$ which enters the metric coefficients is that
    between the events $P$ and $\hat{Q}$; the latter has the same spatial
    coordinates of $Q$ and $\bar{Q}$ but is simultaneous to $P$ with
    respect to the locally barycentric observer $\boldsymbol{u}$ in $P$.
    Here we confuse $\hat{Q}$ with $\bar{Q}$ the error being of the
    order of $1/c^{4}$. In the figure the suffix $(a)$ of the bodies
    are dropped. $\tilde{Q}$ is the position on $S(t)$ of the planet
    which is moving along its world line.}
\end{figure}

Let $x^i(\sigma(t))$ be the spatial coordinates of a general point $P$ on the light
trajectory $\mit\Upsilon$ at the coordinate time $t$ and let $x^i(\tilde\sigma(t))$
be the spatial coordinates of a general  point on the spacetime trajectory of the $a$-th
source, $\tilde\sigma$ being  the parameter along that curve. The metric coefficients
at the point $P$ of the light trajectory are determined by the $a$-th source of gravity
when the latter was located at a point $Q$ of its trajectory at a value  $\tilde\sigma(t')$ 
of its parameter  where $t'$ is the retarded time: $t'=t-|\boldsymbol{\Delta x}(t')|$ where 
$|\boldsymbol {\Delta x}(t')|$ is the modulus of the vector of components $ \Delta x^{i}(t')\equiv x^i(\sigma(t))-
x^{i}(\tilde\sigma(t'))$. We have dropped the suffix $(a)$.
Let the world-line of the given spacetime  source be described by the tangent vector:
\begin{equation}
  \tilde u^{\alpha}(\tilde{\sigma})=\frac{\mathrm{d}x^{\alpha}}
    {\mathrm{d}\tilde{\sigma}}=-(u_{\beta}\tilde u^{\beta})\left(u^{\alpha}+
    \tilde{v}^{\alpha}\right),
     \label{eq:utilde}
\end{equation}
where $\tilde{v}^{\alpha}$ is the $\alpha$-component of the physical 
spatial velocity of the source in the rest frame of the local barycentric 
observer $\boldsymbol{u}$ whose components $u^{\alpha}$ are given 
by (\ref{eq:hatu}); hence we have:
\begin{equation}
  \frac{\mathrm{d}x^i}{\mathrm{d}\tilde{\sigma}}=
    -(u_{\beta}\tilde{u}^{\beta})\tilde{v}^{i}\qquad(i=1,\,2,\,3).
  \label{eq:veloc}
\end{equation}
The distance $r_{_{\cal R}}$ at the retarded time $t'$ along the world-line of the source is given by 
$r_{_{\cal R}}=-\tilde u_\beta \Delta x^\beta$ (see e.g.~\citealp{1999PhRvD..60f124002K}). Since we require 
that our model is accurate to the order of $1/c^3$ then, recalling that the metric potentials 
are at least $1/c^2$, it suffices that the retarded distance $r_{_{\cal R}}$ is determined 
only up to $(1/c)$. From (\ref{eq:utilde}) and recalling that $\Delta x^{i}(t')$ is taken along 
the generators of a light-cone, we have
\begin{equation}
  r_{_{\cal R}}=|\boldsymbol{\Delta x}(t')|-\delta_{ij}\tilde v^i(t')\Delta x^{j}(t').
  \label{eq:retr}
\end{equation}
Before proceeding, let us briefly recall the concept of \textit{spatial} distance. 
Let $r$, say, be such a distance from $P$ to the $a$-th source at $Q$ as it would be  measured by the
barycentric observer $\boldsymbol{u}$ in $P$. From the very definition of a spatial 
distance, the quantity $r$  is the separation between the events $P$ and  $\hat{Q}$ 
the latter being the event of the history of the observer $\boldsymbol{u}$ going trough 
$Q$ which is simultaneous to $P$ with respect to $\boldsymbol{u}$ in $P$ (see figure 
\ref{fig:ret-time}).
\clearpage
\clearpage
From the condition of simultaneity, the coordinates of $\hat{Q}$ are given by:
\begin{eqnarray*}
    x^i_{_{\hat Q}}&=&x^i(\tilde\sigma(t')) \\
    t_{_{\hat Q}}&=&
t+\int^{x^i(\sigma(t))}_{x^i(\tilde\sigma(t'))}\mathrm{e}^{2\psi}h_{0j}(x')\,\mathrm{d}x^{'j}.
\end{eqnarray*}
Here the integral is at least of the order of $1/c^3$ and since it 
would enter terms of an order higher than that we can neglect it 
and confuse $\hat{Q}$ with the event $\bar{Q}$ which is the intersection 
of the world-line of $\boldsymbol{u}$ through $Q$ with the surface $S(t)$.
To first order in $h$ and terming $\tau$ the parameter along the shortest 
space-like path (geodesic) connecting $P$ to $\bar{Q}$ on $S(t)$, the 
required spatial distance reads:
\begin{eqnarray}
  r & = &\int^{\tau(P)}_{\tau(\bar{Q})}\sqrt{{P(u)_{\alpha\beta}\xi^\alpha\xi^\beta}}\,\mathrm{d}\tau\nonumber\\
    & = &\int^{\tau(P)}_{\tau(\bar{Q})}\sqrt{{(\delta_{ij}+h_{ij})\xi^i\xi^j}}\,\mathrm{d}\tau+O(h^2) \label{eq:erre}
\end{eqnarray}
where $\xi^\alpha\equiv\mathrm{d}x^\alpha/\mathrm{d}\tau$ is the unitary 
space-like tangent to the line of integration.  Notice that the integral in (\ref{eq:erre}) contains 
terms of the order $[1/c^0+O(1/c^2)]$ hence, on the basis of what we said before, we can retain only 
the Euclidean part, namely:
\begin{eqnarray}
  r & = & \int^{\tau(P)}_{\tau(\bar{Q})}\sqrt{{\delta_{ij}\xi^i\xi^j}}\,\mathrm{d}\tau+O(1/c^2)\nonumber\\
    & = & \sqrt{\delta_{ij}[x^i(\sigma(t))-x^i(\tilde\sigma(t'))][x^j(\sigma(t))-x^j(\tilde\sigma(t'))]}+O(1/c^2)\nonumber\\
     & \equiv & |\boldsymbol{\Delta x}(t')|+O(1/c^2).
          \label{eq:erreeuclid}
\end{eqnarray}
Here we have made the identifications:
\begin{eqnarray*}
  x^i(\tau(\bar Q)) & = & x^i(\tilde\sigma(t')) \\
  x^i(\tau(P)) & = & x^i(\sigma(t)).
\end{eqnarray*}

From (\ref{eq:erreeuclid}) and (\ref{eq:retr}) we can write the metric coefficients (\ref{eq:hform}) as
{\arraycolsep=2pt
\begin{equation}
  \left\{
  \begin{array}{rcl}
    h_{00}^{(a)}&=&\left(\displaystyle{\frac{2G\mathcal{M}^{(a)}}{c^{2}r^{(a)}}}\right)
      \left(1+\mathbf{v}^{(a)}\cdot\hat{\mathbf{n}}^{(a)}\right)+O\left(\displaystyle{\frac{1}{c^{4}}}\right) \\[15pt]
    h_{jk}^{(a)}&=&\left(\displaystyle{\frac{2G\mathcal{M}^{(a)}}{c^{2}r^{(a)}}}\right)\left(1+
      \mathbf{v}^{(a)}\cdot\hat{\mathbf{n}}^{(a)}\right)\delta_{jk}+O\left(\displaystyle{\frac{1}{c^{4}}}\right) \\[15pt]
    h_{0j}^{(a)}&=&\displaystyle{-\frac{2w_{j}^{(a)}}{c^{3}}}
                   +O\left(\displaystyle{\frac{1}{c^{4}}}\right),
  \end{array}
  \right.
\label{eq:h-approx}
\end{equation}
}where $\mathbf{n}^{(a)}=\mathbf{r}^{(a)}/r^{(a)}$, all quantities here being calculated at 
the retarded time $t'$. Having defined as retarded distance the quantity $r_{\cal R}$, we shall 
term {\textit {reduced}} distance the quantity $r$ as in  (\ref{eq:erreeuclid}). 

It is convenient, for computational purposes, to express the distance $r$ in terms of quantities
defined all at the same time $t$ of metric determination. To this purpose few more steps are needed
for the correct identification of the retarded time corrections. Integrating (\ref{eq:veloc}) from $\tilde{\sigma}(t')$ to $\tilde{\sigma}(t)$ which correspond respectively to the points $Q$ and 
$\tilde Q$ as shown in figure~\ref{fig:ret-time}
\begin{equation}
  x^{i}(\tilde{\sigma}(t'))=x^{i}(\tilde{\sigma}(t))+
    \int^{\tilde{\sigma}(t)}_{\tilde{\sigma}(t')}
    (u_{\beta}\tilde{u}^{\beta})\tilde{v}^{i}(\tilde{\sigma})\,\mathrm{d}\tilde{\sigma}\cdots
  \label{eq:coorplanet2}
\end{equation}
the integral being taken along the world-line of the gravity source. From this it follows:
\begin{equation}
  x^{i}(\sigma(t))-x^{i}(\tilde{\sigma}(t'))=
    x^{i}(\sigma(t))-x^{i}(\tilde{\sigma}(t))-
    \int^{\tilde{\sigma}(t)}_{\tilde{\sigma}(t')}
    (u_{\beta}\tilde{u}^{\beta})\tilde{v}^{i}(\tilde{\sigma})\,\mathrm{d}\tilde\sigma+\cdots.
  \label{eq:erre2}
\end{equation}
Notice that $x^{i}(\tilde{\sigma}(t))$ are the spatial coordinates 
of the gravity source at the time $t$ in the BCRS; these are supposed 
to be known; also the components of its spatial velocity $\tilde{v}^{i}(\tilde{\sigma}(t))$ 
with respect to the local barycentric observer at time $t$ are assumed to be known.
From (\ref{eq:erre}) and (\ref{eq:erre2}) we then have:
\begin{eqnarray}
  r= & \left\{\vphantom{\displaystyle{\int^{\tilde{\sigma}(t)}_{\tilde{\sigma}(t')}}}\delta_{ij}\right. &
       \left[x^{i}(\sigma(t))-x^{i}(\tilde{\sigma}(t))-
       \int^{\tilde{\sigma}(t)}_{\tilde{\sigma}(t')}(u_{\beta}\tilde{u}^{\beta})\tilde{v}^{i}
       (\tilde{\sigma})\,\mathrm{d}\tilde{\sigma}\cdots\right]\times\nonumber\\
     & & \left.\left[x^{j}(\sigma(t))-x^{j}(\tilde{\sigma}(t))-
       \int^{\tilde{\sigma}(t)}_{\tilde{\sigma}(t')}(u_{\beta}\tilde{u}^{\beta})\tilde{v}^{j}
       (\tilde{\sigma})\,\mathrm{d}\tilde{\sigma}+\cdots\right]\right\}^{1/2}+O\left(\frac{1}{{c^2}}\right)
  \label{eq:rquadratic}
\end{eqnarray}
Setting
\begin{equation}
  \Delta x^{i}(t)=x^{i}(\sigma(t))-x^{i}(\tilde{\sigma}(t)).
  \label{eq:deltax}
\end{equation}
and recalling that
\begin{equation}
  -(u^{\alpha}\tilde{u}_{\alpha})=(1-\tilde{v}^{2})^{-1/2}
  \label{eq:gammasource}
\end{equation}
$\tilde{v}$ being the (modulus of the) spatial velocity of the gravity source 
with respect to the local barycentric observer, equation (\ref{eq:rquadratic}) 
writes as:
\begin{eqnarray}
  r= & \left\{\vphantom{\displaystyle{\int^{\tilde{\sigma}(t)}_{\tilde{\sigma}(t')}}}\delta_{ij}\right. &
       \left[\Delta x^{i}(t)+\int^{\tilde{\sigma}(t)}_{\tilde{\sigma}(t')}
       \left(1+\frac{1}{2}\tilde{v}^{2}\right)\tilde{v}^{i}(\tilde{\sigma})\,\mathrm{d}\tilde{\sigma}\cdots\right]
       \times\nonumber\\
     & &\left.\left[\Delta x^{j}(t)+\int^{\tilde{\sigma}(t)}_{\tilde{\sigma}(t')}
       \left(1+\frac{1}{2}\tilde{v}^{2}\right)\tilde{v}^{j}(\tilde{\sigma})\,\mathrm{d}\tilde{\sigma}
       +\cdots\right]\right\}^{1/2}
  \label{eq:newrquadratic}
\end{eqnarray}
which can be written  as follows:
\begin{equation}
  r=\left[r_{0}^{2}+2\delta_{ij}\Delta x^{i}(t)\int^{\tilde{\sigma}(t)}_{\tilde{\sigma}(t')}
    \tilde{v}^{j}\,\mathrm{d}\tilde{\sigma}+\delta_{ij}\int^{\tilde{\sigma}(t)}_{\tilde{\sigma}(t')}
    \tilde{v}^{i}\,\mathrm{d}\tilde{\sigma}\int^{\tilde{\sigma}(t)}_{\tilde{\sigma}(t')}
    \tilde{v}^{j}\,\mathrm{d}\tilde{\sigma}\right]^{1/2}+O\left(\frac{1}{{c^2}}\right)
  \label{eq:rquadraticfinal}
\end{equation}
 where at any $t$
\begin{equation}
  r_0=\sqrt{\delta_{ij}\Delta x^i(t)\Delta x^j(t)}.
\end{equation}
For the bodies of the Solar System the quantities 
$\int^{\tilde{\sigma}(t)}_{\tilde{\sigma}(t')}\tilde{v}^{j}(\tilde{\sigma})\,\mathrm{d}\tilde{\sigma}$ 
are  always much smaller then $r_{0}$; even in the extreme case of a light 
ray skimming the surface of a planet the integral goes to zero while $r_0$ 
remains of the order of the body's radius. Thus the following Taylor expansion 
is justified. Remembering that $(1+x)^{1/2}\approx1+(1/2)x-(1/8)x^{2}+\cdots$,
then to the given order we have from equation (\ref{eq:rquadraticfinal}):
\begin{eqnarray}
r & = & r_{0}\left\{\vphantom{\left[\int_{\tilde{\sigma}(t')}^{\tilde{\sigma}(t)}\right]^{2}} 1+\frac{1}{r_{0}^{2}}\left[\delta_{ij}\Delta x^{i}(t)
\int_{\tilde{\sigma}(t')}^{\tilde{\sigma}(t)}\tilde{v}^{j}(\tilde{\sigma})
\,\mathrm{d}\tilde{\sigma}+\frac{1}{2}\delta_{ij}
\int_{\tilde{\sigma}(t')}^{\tilde{\sigma}(t)}\tilde{v}^{i}(\tilde{\sigma})
\,\mathrm{d}\tilde{\sigma}
\int_{\tilde{\sigma}(t')}^{\tilde{\sigma}(t)}\tilde{v}^{j}(\tilde{\sigma})
\,\mathrm{d}\tilde{\sigma}\right]\right.\nonumber \\
 & - & \left.\frac{1}{8r_{0}^{4}}\left[\delta_{ij}\Delta x^{i}(t)
 \int_{\tilde{\sigma}(t')}^{\tilde{\sigma}(t)}\tilde{v}^{j}(\tilde{\sigma})
 \,\mathrm{d}\tilde{\sigma}+\frac{1}{2}\delta_{ij}
 \int_{\tilde{\sigma}(t')}^{\tilde{\sigma}(t)}\tilde{v}^{i}(\tilde{\sigma})
 \,\mathrm{d}\tilde{\sigma}
 \int_{\tilde{\sigma}(t')}^{\tilde{\sigma}(t)}\tilde{v}^{j}(\tilde{\sigma})
 \,\mathrm{d}\tilde{\sigma}\right]^{2}\right\} \nonumber \\
 & + & O\left(\frac{1}{c^{2}}\right).
\label{eq:rfinal}
\end{eqnarray}
Evidently, the $1/c^{2}$ contribution of the relative velocity of
the gravitating sources to the retarded time correction can be neglected.
The integrations in (\ref{eq:rfinal}) contain the unknown $r$ implicitly
in $\tilde{\sigma}(t')$ and this makes the calculations rather cumbersome,
however order of magnitude considerations allow one to identify the proper-time 
$\tilde{\sigma}(t)$ with the coordinate time $t$ so achieving a considerable 
simplification. From (\ref{eq:utilde}), in fact, we have
\begin{equation}
  \tilde{u}^{0}=\frac{\mathrm{d}t}{\mathrm{d}\tilde{\sigma}}=
    \tilde{\gamma}\left(u^{0}+\tilde{v}^{0}\right)
  \label{eq:utildezero}
\end{equation}
where $\tilde{\gamma}=-\tilde{u}_{\sigma}u^{\sigma}=(1-\tilde{v}^{2})^{-1/2}$
is the Lorentz factor of the gravity source relative to the local
barycentric observer. Recalling (\ref{eq:metric}) and (\ref{eq:hatu})
we easily get:
\begin{eqnarray}
u^{0} & = & 1+\frac{1}{2}h_{00}+O\left(\frac{1}{c^{4}}\right)\nonumber \\
\tilde{v}^{0} & = & O\left(\frac{1}{c^{4}}\right)
\end{eqnarray}
the latter relation arising from the condition $\tilde{v}_{\alpha}u^{\alpha}=0$.
Thus equation (\ref{eq:utildezero}) becomes:
\begin{equation}
\frac{\mathrm{d}t}{\mathrm{d}\tilde{\sigma}}=
1+\frac{1}{2}h_{00}+\frac{1}{2}\tilde{v}^{2}+O\left(\frac{1}{c^{4}}\right)
\end{equation}
which leads to
\begin{equation}
\mathrm{d}\tilde{\sigma}=\mathrm{d}t+O\left(\frac{1}{c^{2}}\right).
\end{equation}
A similar argument applies to the integration variable $\sigma$
entering the master equations (\ref{eq:masterequation1}) and (\ref{eq:masterequation2})
and so, to the required order, we can calculate the individual terms
in (\ref{eq:masterequation1}) and (\ref{eq:masterequation2}) and
in (\ref{eq:rfinal}) confusing the barycentric and the gravity source
proper-times with the coordinate time $t$. From the above considerations
and setting $\Delta\tilde{x}^{i}(t,t')\equiv\tilde{x}^{i}(t)-
\tilde{x}^{i}(t')=\int_{t'}^{t}\tilde{v}^{i}(t)\,\mathrm{d}t+O(1/c^{2})$ with $\tilde x^i(t)
\equiv x^i(\tilde\sigma(t))$,
equation (\ref{eq:rfinal}) becomes:
\begin{eqnarray}
r & = & r_{0}\left\{\vphantom{\left[\frac{1}{2}\right]^{2}}1+\frac{1}{r_{0}^{2}}\left[\delta_{ij}\Delta x^{i}(t)\Delta\tilde{x}^{j}(t,t')+\frac{1}{2}\delta_{ij}
\Delta\tilde{x}^{i}(t,t')\Delta\tilde{x}^{j}(t,t')\right]\right.\nonumber \\
 & - & \left.\frac{1}{8r_{0}^{4}}\left[\delta_{ij}\Delta x^{i}(t)
 \Delta\tilde{x}^{j}(t,t')+\frac{1}{2}\delta_{ij}\Delta\tilde{x}^{i}(t,t')
 \Delta\tilde{x}^{j}(t,t')\right]^{2}\right\} +O\left(\frac{1}{c^{2}}\right).
\label{eq:rfinalt}
\end{eqnarray}

It is clear that, being $\tilde{v}^{j}(t)$ a bounded function, the
integrals in (\ref{eq:rfinalt}) remain finite when $r_{0}\rightarrow\infty$,
hence in that limit we have that $r\rightarrow r_{0}$ as expected.
Let us now define $\mathbf{r}_{0}=\{\Delta x^{i}(t)\}$ and $\tilde{\mathbf{r}}=\{\Delta\tilde{x}^{i}(t,t')\}$
with $i=1,\,2,\,3$; since the orbits of the spacetime sources are
bounded then $\tilde{\mathbf{r}}$ is finite while $\mathbf{r}_{0}$
can grow to infinity, then we retain only terms of the order of $(\tilde{\mathbf{r}}/\mathbf{r}_{0})^{2}$.
Equation (\ref{eq:rfinalt}) can then be cast in vectorial form:
\begin{equation}
r=r_{0}\left[1+\frac{1}{r_{0}^{2}}\left(\mathbf{r}_{0}\cdot\tilde{\mathbf{r}}+
\frac{1}{2}\tilde{\mathbf{r}}\cdot\tilde{\mathbf{r}}\right)-
\frac{1}{8r_{0}^{4}}\left(\mathbf{r}_{0}\cdot\tilde{\mathbf{r}}\right)^{2}+
O\left(\left(\frac{||\tilde{\mathbf{r}}||}{||\mathbf{r}_{0}||}\right)^{3}\right)\right]+
O\left(\frac{1}{c^{2}}\right).
\label{eq:rvector}
\end{equation}
It should be stressed here that (\ref{eq:rvector}) is an implicit
equation since $\tilde{\mathbf{r}}$ is function of $r$ in its turn.

\section{The boundary conditions\label{sec:boundary-conditions}}
The differential equations (\ref{eq:masterequation1}) and (\ref{eq:masterequation2})
are of the general form:
\begin{equation}
\frac{\mathrm{d}\bar{\ell}^{\alpha}}{\mathrm{d}\sigma}=
\mathcal{F}^{\alpha}(\partial_{\beta}h(x,y,z,t),\bar{\ell}^{i}(\sigma(x)))
\label{eq:lightray}
\end{equation}
where $\mathcal{F}^{\alpha}$ are real, non singular, smooth functions
of their arguments. A general solution of (\ref{eq:lightray}) is
\begin{equation}
\bar{\ell}^{\alpha}=\bar{\ell}^{\alpha}(\sigma,\bar{\ell}_{(0)}^{k})
\end{equation}
where $\bar{\ell}_{(0)}^{k}$ are the components of the vector $\bar{\boldsymbol{\ell}}$
at the observation and represent the boundary values that we need
to fix in order to integrate (\ref{eq:lightray}). These boundary
conditions can only be expressed in terms of the satellite observations.
In the case of Gaia the observables are the angles that the incoming
light ray forms with the spatial axes of the satellite attitude triad.
The latter is a set of three orthonormal space-like vectors which
are comoving with the satellite and define its rest frame; their coordinate
components are denoted as $\{ E_{\hat{a}}^{\alpha}\}_{\hat{a}=1,2,3}$
and satisfy the condition $E_{\hat{a}}^{\alpha}E_{\alpha\hat{b}}=\delta_{\hat{a}\hat{b}}$.
The observables are given by:
\begin{equation}
\cos\psi_{(E_{\hat{a}},\bar{\ell})}\equiv\mathbf{e}_{\hat{a}}=
\frac{P(u')_{\alpha\beta}k^{\alpha}E_{\hat{a}}^{\beta}}
{(P(u')_{\alpha\beta}k^{\alpha}k^{\beta})^{1/2}};
\label{eq:cos}
\end{equation}
$P(u')_{\alpha\beta}$ is the operator which projects into the satellite's
rest-frame, namely:
\begin{equation}
P(u')_{\alpha\beta}=g_{\alpha\beta}+u'_{\alpha}u'_{\beta}
\end{equation}
and $\boldsymbol{u}'$ is the time-like and unitary satellite four
velocity. The latter is given by
\begin{equation}
\boldsymbol{u}'=T_{s}(\boldsymbol{\partial}_{t}+
\beta_{1}\boldsymbol{\partial}_{x}+\beta_{2}\boldsymbol{\partial}_{y}+
\beta_{3}\boldsymbol{\partial}_{z}),
\label{eq:us}
\end{equation}
where $\boldsymbol{\partial}_{\alpha}$'s are the coordinate basis
vectors relative to the barycentric celestial reference system, $\beta_{i}$
are the \textit{coordinate} components of the satellite three-velocity
with respect to the barycenter of the Solar System. Here the subscripts
refer to contravariant components not to confuse them with power indeces.
It is important to stress that the modulus of $\beta_{i}$ is \textit{not}
the physical velocity of the satellite with respect to the local barycentric
observer $\boldsymbol{u}$, nor the $\beta_{i}$ are its physical
components. The physical velocity instead, namely the quantity which
would be measured, is given by the modulus of the four vector with
components:
\begin{equation}
\nu^{\alpha}=\frac{1}{\gamma}\left(u'^{\alpha}-\gamma u^{\alpha}\right)
\label{eq:ni}
\end{equation}
where $\gamma$ is the instantaneous satellite's Lorentz factor. Form (\ref{eq:ni}) it follows: 
\begin{eqnarray}
\nu^{0} & = & -\frac{1}{2}U\beta^{2}=O(1/c^{4})\nonumber \\
\nu^{i} & = & (1+U)\beta_{i}+O(1/c^{4})
\label{eq:ni3}
\end{eqnarray}
where $2U=h_{00}$ $\beta^{2}=\beta_{1}^{2}+\beta_{2}^{2}+\beta_{3}^{2}$; hence
$\nu^{i}$ coincides with $\beta_{i}$ only to the order of $1/c^{2}$.
Finally we recall that $u'^{\alpha}u'_{\alpha}=-1$ and this relation
fixes the time factor of $\boldsymbol{u}'$ as $T_{s}=1+\frac{1}{2}(h_{00}+\beta^{2})$.
From (\ref{eq:xidot}) and (\ref{eq:us}), equation (\ref{eq:cos})
can be written more explicitely as:
\begin{equation}
\cos\psi_{(E_{\hat{a}},\bar{\ell})}\equiv\mathbf{e}_{\hat{a}}=
\frac{P(u')_{\alpha\beta}(\bar{\ell}_{(0)}^{\alpha}+
u^{\alpha}){E_{\hat{a}}}^{\beta}}{[u'_{\rho}(\bar{\ell}_{(0)}^{\rho}+
u^{\rho})]^{1/2}}=-\frac{(\bar{\ell}_{(0)}^{\beta}-
\nu^{\beta})E_{\hat{a}\beta}}{\gamma\nu_{\alpha}\bar{\ell}_{(0)}^{\alpha}-\gamma}
\label{eq:ea}
\end{equation}
all terms being calculated at the observation time. We easily see
that all quantities contained in (\ref{eq:ea}) are known except $\bar{\ell}_{(0)}^{i}$
which obviously are the unknown boundary conditions, as stated. In
order to deduce them correctly one needs the components of the satellite
attitude triad. They have been deduced in explicit form in \citet{2003CQGra..20.4695B}
(see equations (4.17) and Appendix B of that paper). Here we shall cast them in a form easier to 
handle in direct calculations.

The satellite is expected to rotate with an angular velocity $\omega_{r}$
about an axis (its $\hat{x}$-axis) which forms an angle $\alpha$
with respect to the instantaneous local direction to the Sun. The
spin axis then precesses%
\footnote{In the case of Gaia the satellite will make one turn every $\sim6$ hours
with a precession period of $\sim70$ days and a precession angle $\alpha$
of $\sim45^{\circ}$.%
} about the direction to the Sun with an angular velocity $\omega_{p}$.
The attitude triad will then depend on this parameter specification
as follows \citep{2003CQGra..20.4695B}:
\begin{eqnarray}
  E_{\hat{1}}^{\alpha} & = & \cos\alpha\underset{\mathrm{bs}}{\lambda_{\hat{1}}^{\alpha}}-
    \sin\alpha\cos(\omega_{p}t)\underset{\mathrm{bs}}{\lambda_{\hat{2}}^{\alpha}}-
    \sin\alpha\cos(\omega_{p}t)\underset{\mathrm{bs}}{\lambda_{\hat{3}}^{\alpha}}\label{eq:E1}\\
  E_{\hat{2}}^{\alpha} & = & -\sin\alpha\sin(\omega_{r}t)
    \underset{\mathrm{bs}}{\lambda_{\hat{1}}^{\alpha}}+\nonumber \\
   &  & +[\cos(\omega_{r}t)\cos(\omega_{p}t)-
    \sin(\omega_{r}t)\sin(\omega_{p}t)\cos\alpha]
    \underset{\mathrm{bs}}{\lambda_{\hat{2}}^{\alpha}}\label{eq:E2}\\
   &  & +[\cos(\omega_{r}t)\sin(\omega_{p}t)+
    \sin(\omega_{r}t)\cos(\omega_{p}t)\cos\alpha]
    \underset{\mathrm{bs}}{\lambda_{\hat{3}}^{\alpha}}\nonumber \\
  E_{\hat{3}}^{\alpha} & = & -\sin\alpha\cos(\omega_{r}t)
    \underset{\mathrm{bs}}{\lambda_{\hat{1}}^{\alpha}}\nonumber \\
   &  & -[\sin(\omega_{r}t)\cos(\omega_{p}t)+
    \cos(\omega_{r}t)\sin(\omega_{p}t)\cos\alpha]
    \underset{\mathrm{bs}}{\lambda_{\hat{2}}^{\alpha}}\label{eq:E3}\\
   &  & +[-\sin(\omega_{r}t)\sin(\omega_{p}t)+
    \cos(\omega_{r}t)\cos(\omega_{p}t)\cos\alpha]
    \underset{\mathrm{bs}}{\lambda_{\hat{3}}^{\alpha}}\nonumber
\end{eqnarray}
where $\alpha$ takes values as ($t,\, x,\, y,\, z$) and 
$\{\underset{\mathrm{bs}}{\boldsymbol{\lambda}_{\hat{a}}}\}$
is the Lorentz boosted triad adapted to the satellite whose components are given 
in explicit form in the appendix~\ref{app:lb}.

\subsection{The spatial velocity of the Gaia satellite}
To make the boundary conditions complete we need an operational definition
of the satellite's spatial physical velocity $\nu^{i}$. The satellite
trajectory will be close to that of the outer Lagrangian point $L_{2}$
relative to the Earth-Sun system. The world lines of $L_{2}$ and
Gaia never intersect because Gaia moves around $L_{2}$ in a halo
type orbit. Hence to define Gaia's spatial velocity with respect to
$L_{2}$ we have first to fix a coordinate frame comoving with $L_{2}$.
Denoting the coordinates with respect to $L_{2}$ with a bar, we define:\[
\bar{t}=t\qquad\bar{x}_{_{G}}=x_{_{G}}-x_{_{L_{2}}}\]
 hence the four-velocity of Gaia in the coordinate frame of $L_{2}$
reads:
\[
\bar{u}_{_{G}}^{\alpha}=\bar{u}_{_{G}}^{0}\left(\delta_{0}^{\alpha}+
\bar{\beta}_{_{G}}^{i}\delta_{i}^{\alpha}\right)
\]
where
\[
\bar{\beta}_{_{G}}^{i}=\frac{\mathrm{d}\bar{x}_{_{G}}^{i}}{\mathrm{d}t}
\]
are the coordinate components of Gaia's spatial velocity with respect to $L_{2}$.

Let us identify what are the unknowns and what is known. Our task
is to express the formers in terms of the latters. The \textit{unknowns}
are the components of the spatial velocity of Gaia with respect to
the local barycentric observer, namely $\nu^{i}$. The \textit{known}
quantities are: \emph{(i)} the components of the coordinate spatial
velocity of $L_{2}$ with respect to the local barycentric observer
namely $\beta_{L_{2}}^{i}=\mathrm{d}x_{L_{2}}^{i}/\mathrm{d}t$;
\emph{(ii)} the components of the spatial velocity of Gaia with respect
to $L_{2}$, namely $\bar{\beta}_{G}$; \emph{(iii)} the metric
coefficients at the position of Gaia. 

Let us express $\nu^{i}$ in terms of the known quantities. From (\ref{eq:ni})
it follows that
\begin{equation}
\nu^{i}=\beta^{i}\left(1+\frac{1}{2}h_{00}+O(1/c^{4})\right)
\label{eq:ni1}
\end{equation}
hence, recalling that $\beta^{i}=\mathrm{d}x_{G}^{i}/\mathrm{d}t$,
we have
\begin{equation}
\beta^{i}=\frac{\mathrm{d}\bar{x}_{G}^{i}}{\mathrm{d}t}+
\frac{\mathrm{d}x_{L_{2}}^{i}}{\mathrm{d}t}=\bar{\beta}_{G}^{i}+\beta_{L_{2}}^{i}
\end{equation}
namely
\begin{equation}
\nu^{i}=\bar{\beta}_{G}^{i}\left(1+\frac{1}{2}h_{00}+O(1/c^{4})\right)+\nu_{L_{2}}^{i}.
\label{eq:ni2}
\end{equation}
In (\ref{eq:ni1}) and (\ref{eq:ni2}) the gravitational potential
$h_{00}$ is calculated at the position of the satellite.

\section{Testing RAMOD4\label{sec:Tests}}
The most efficient way to test RAMOD4 is to compare
the results of the integration of its equations of motion with those
of RAMOD3. We expect that the differences between the two models are
of the order of $5\cdot10^{-11}$~rad {\em i.e.} $\sim10\mu$as (see 
section 7 of \citealp{2004ApJ...607..580D}).
Obviously the practical implementation of these equations requires
the precise specification of some of its ingredients. First of all an 
explicit form of the metric (\ref{eq:h-approx}) should be chosen;
in this context and to the prescribed order of $1/c^3$ we chose, as stated early, a 
solution of Einstein's equations based on the Li\'enard-Wiechert tensor 
potentials in terms of which, we recall, the post-Minkowskian metric perurbations 
take the form
\begin{eqnarray}
  h_{00} & = & \sum_{a}\frac{2G\mathcal{M}^{(a)}}{c^{2}r^{(a)}}
    \left(1+\mathbf{v}^{(a)}\cdot\hat{\mathbf{n}}^{(a)}\right)+O\left(\frac{1}{c^{4}}\right)\nonumber \\
  h_{0i} & = & \sum_{a}-\frac{4G\mathcal{M}^{(a)}}{c^{3}r^{(a)}}\tilde{v}_{i}^{(a)}
    =-2h_{00}\frac{\tilde{v}_{i}}{c}+O\left(\frac{1}{c^{4}}\right)\label{eq:metric-explicit}\\
  h_{ij} & = & \sum_{a}\frac{2G\mathcal{M}^{(a)}}{c^{2}r^{(a)}}
    \left(1+\mathbf{v}^{(a)}\cdot\hat{\mathbf{n}}^{(a)}\right)\delta_{ij}=
    h_{00}\delta_{ij}+O\left(\frac{1}{c^{4}}\right).\nonumber 
\end{eqnarray}
where the potential $w_{j}^{(a)}$ due to the relative velocities of the bodies
is equal to
\[
w_{j}^{(a)}=-\frac{4G\mathcal{M}^{(a)}}{c^{3}r^{(a)}}\tilde{v}_{j}^{(a)}+O\left(\frac{1}{c^{4}}\right).
\]
The coordinate form for the master equations obtained by using the explicit metric 
components (\ref{eq:metric-explicit}) in (\ref{eq:masterequation2}) 
is reported in appendix~\ref{app:me}. The reduced distance $r^{(a)}$ which
enters equations (\ref{eq:metric-explicit}) will be treated separately
in subsection~\ref{sub:implementing-reduced}.

The boundary conditions which are needed to integrate (\ref{eq:masterequation2})
can be specified only after the tetrad describing the motion of the
observer is given explicitely. Our goal is to compare this new
model with RAMOD3, hence a natural choice is a tetrad
that makes the measurements compatible with those described in \citet{2004ApJ...607..580D},
that is, a phase-locked tetrad associated to an observer moving on
a circular orbit around the barycenter of the Solar System on the
plane $z=0$.

The four-velocity $u'^{\alpha}$ of this observer writes
\begin{equation}
  u'^{\alpha}=\mathrm{e}^{\psi'}\left[\delta_{0}^{\alpha}-
    \omega\left(y_{\mathrm{s}}-y_{\odot}\right)\delta_{x}^{\alpha}+
    \omega\left(x_{\mathrm{s}}-x_{\odot}\right)\delta_{y}^{\alpha}\right],
  \label{eq:u-satellite}
\end{equation}
where $\omega$ is the angular velocity of the observer, $x_{\odot}$
and $y_{\odot}$ are the barycentric position of the Sun, and $\mathrm{e}^{\psi'}$
is a normalization factor required by the condition $u'^{\alpha}u'_{\alpha}=-1$.
Since in RAMOD4 the only non-vanishing terms of the metric are $g_{|\alpha|\alpha}$
and $g_{0i}$, this factor is
\[ 
  \mathrm{e}^{\psi'}=\left\{ -\left[g_{00}-2\omega\left(y_{\mathrm{s}}-y_{\odot}\right)
    g_{0x}+2\omega\left(x_{\mathrm{s}}-x_{\odot}\right)g_{0y}+
    \omega^{2}\left(y_{\mathrm{s}}-y_{\odot}\right)^{2}g_{xx}+
    \omega^{2}\left(x_{\mathrm{s}}-x_{\odot}\right)^{2}g_{yy}\right]\right\} ^{-1/2}.
\]

The explicit expression of the components of the tetrad spatial axes
can be deduced by the components of the tetrad $\left\{
\underset{\mathrm{bs}}{\lambda_{\hat{a}}}\right\} $ given in 
eqs.(\ref{eq:Lbs1t}-\ref{eq:sun-coord}) setting $\theta=0$ and 
substituting the velocity components of the observer with the following 
expressions
\begin{equation}
  \begin{array}{rcl}
    \beta_{1} & = & -\omega\left(y_{\mathrm{s}}-y_{\odot}\right)\mathrm{e}^{\psi'}\\
    \beta_{2} & = & \omega\left(x_{\mathrm{s}}-x_{\odot}\right)\mathrm{e}^{\psi'}\\
    \beta_{3} & = & 0
  \end{array}
  \label{eq:beta-comp}
\end{equation}
which describe, as stated, a spatially circular orbit.

From (\ref{eq:beta-comp}) and following the conventions in \citet{2003CQGra..20.4695B},
the tetrad component $\underset{\mathrm{bs}}{\lambda}_{\hat{1}}^{0}$
become
\[
\begin{array}{rcl}
  \underset{\mathrm{bs}}{\lambda_{\hat{1}}^{0}} & = & \left(\frac{3}{2}h_{00}+\frac{1}{2}\omega^{2}R^{2}
    \mathrm{e}^{2\psi'}+1\right)\left[-\omega\left(y_{\mathrm{s}}-y_{\odot}\right)
    \mathrm{e}^{\psi'}\cos\phi_{\mathrm{s}}+
    \omega\left(x_{\mathrm{s}}-x_{\odot}\right)\mathrm{e}^{\psi'}\sin\phi_{\mathrm{s}}\right]+\\
   & + & \left(h_{0x}\cos\phi_{\mathrm{s}}+h_{0y}\sin\phi_{\mathrm{s}}\right),
\end{array}
\]
where $R$ is the radius of the circular orbit of the observer. Since
$\left(x_{\mathrm{s}}-x_{\odot}\right)=R\cos\phi_{\mathrm{s}}$ and
$\left(y_{\mathrm{s}}-y_{\odot}\right)=R\sin\phi_{\mathrm{s}}$, the
above expression simplifies as
\begin{eqnarray*}
  \underset{\mathrm{bs}}{\lambda_{\hat{1}}^{0}} & = & \left(\frac{3}{2}h_{00}+\frac{1}{2}\omega^{2}R^{2}
    \mathrm{e}^{2\psi'}+1\right)\frac{\omega\mathrm{e}^{\psi'}}{R}
    \left[-\left(x_{\mathrm{s}}-x_{\odot}\right)\left(y_{\mathrm{s}}-y_{\odot}\right)+
    \left(x_{\mathrm{s}}-x_{\odot}\right)\left(y_{\mathrm{s}}-y_{\odot}\right)\right]+\\
   & + & \frac{\left[h_{0x}\left(x_{\mathrm{s}}-
    x_{\odot}\right)+h_{0y}\left(y_{\mathrm{s}}-y_{\odot}\right)\right]}{R}\\
   & = & \frac{1}{R}\left[h_{0x}\left(x_{\mathrm{s}}-
    x_{\odot}\right)+h_{0y}\left(y_{\mathrm{s}}-y_{\odot}\right)\right].
\end{eqnarray*}
Similarly, the other components of $\underset{\mathrm{bs}}{\lambda_{\hat{1}}}$
turn out to be
\begin{eqnarray}
  \underset{\mathrm{bs}}{\lambda_{\hat{1}}^{x}} & = & \frac{\left(x_{\mathrm{s}}-
    x_{\odot}\right)}{R}\left(1-\frac{h_{00}}{2}\right)\nonumber \\
  \underset{\mathrm{bs}}{\lambda_{\hat{1}}^{y}} & = & \frac{1}{R}\left[\left(x_{\mathrm{s}}-x_{\odot}
    \right)-\frac{1}{2}h_{00}\left(y_{\mathrm{s}}-y_{\odot}\right)\right]\label{eq:comp-tetrad-1}\\
  \underset{\mathrm{bs}}{\lambda_{\hat{1}}^{z}} & = & 0,\nonumber
\end{eqnarray}
while the remaining axes are
\begin{equation}
  \begin{array}{rcl}
    \underset{\mathrm{bs}}{\lambda_{\hat{2}}^{0}} & = & \left(\frac{3}{2}h_{00}+
      \frac{1}{2}\omega^{2}R^{2}\mathrm{e}^{2\psi'}+1\right)\omega R\mathrm{e}^{\psi'}+
      \frac{-h_{0x}\left(y_{\mathrm{s}}-y_{\odot}\right)+h_{0y}\left(x_{\mathrm{s}}-x_{\odot}\right)}{R}\\
    \underset{\mathrm{bs}}{\lambda_{\hat{2}}^{x}} & = & -\frac{1}{2}\omega^{2}R\left(y_{\mathrm{s}}-
      y_{\odot}\right)\mathrm{e}^{2\psi'}-\frac{\left(y_{\mathrm{s}}-
      y_{\odot}\right)}{R}\left(1-\frac{h_{00}}{2}\right)\\
    \underset{\mathrm{bs}}{\lambda_{\hat{2}}^{y}} & = & \frac{1}{2}\omega^{2}R\left(x_{\mathrm{s}}-
      x_{\odot}\right)\mathrm{e}^{2\psi'}+\frac{\left(x_{\mathrm{s}}-
      x_{\odot}\right)}{R}\left(1-\frac{h_{00}}{2}\right)\\
    \underset{\mathrm{bs}}{\lambda_{\hat{2}}^{z}} & = & 0,\end{array}\label{eq:comp-tetrad-2}
\end{equation}
and
\begin{equation}
  \begin{array}{rcl}
    \underset{\mathrm{bs}}{\lambda_{\hat{3}}^{0}} & = & h_{0z}\\
    \underset{\mathrm{bs}}{\lambda_{\hat{3}}^{x}} & = & 0\\
    \underset{\mathrm{bs}}{\lambda_{\hat{3}}^{y}} & = & 0\\
    \underset{\mathrm{bs}}{\lambda_{\hat{3}}^{z}} & = & 1-\frac{h_{00}}{2}.
  \end{array}
  \label{eq:comp-tetrad-3}
\end{equation}

Finally, given these tetrad components, one can fix the boundary conditions
$\bar{\ell}_{(0)}^{i}$ needed for the integration of the equations
of motion; inverting eq.(\ref{eq:ea}) and to the $1/c^{3}$ order,
they are given by:
\[
  \bar{\ell}_{(0)}^{i}=\frac{N^{i}}{D}\qquad i=x,y,z
\]
where
\begin{eqnarray*}
  D & = & -\mathbf{e}_{\hat{2}}\beta\left(1+4U+\frac{1}{2}\beta^{2}\right)+
    \left(1+2U+\frac{1}{2}\beta^{2}\right)\\
  N^{x} & = & \frac{y}{R}\left[\mathbf{e}_{\hat{2}}\left(1+U+\frac{1}{2}\beta^{2}\right)-
    \beta\left(1+3U+\frac{1}{2}\beta^{2}\right)\right]-
    \frac{x}{R}\mathbf{e}_{\hat{1}}\left(1+U\right)\\
  N^{y} & = & \frac{x}{R}\left[-\mathbf{e}_{\hat{2}}\left(1+U+\frac{1}{2}\beta^{2}\right)+
    \beta\left(1+3U+\frac{1}{2}\beta^{2}\right)\right]-\frac{y}{R}\mathbf{e}_{\hat{1}}\left(1+U\right)\\
  N^{z} & = & -\mathbf{e}_{\hat{3}}\left(1+U\right).
\end{eqnarray*}

\subsection{Implementing the reduced distance\label{sub:implementing-reduced}}
Let us recall that all the metric coefficients contain the reduced distance of 
the perturbing bodies, so its practical implementation into the formulae for 
the tests requires to make explicit assumptions on the ephemeris. This was not 
needed in RAMOD3 since the metric is stationary  in that model. Moreover, in 
comparing RAMOD3 to RAMOD4, we are at liberty to choose the ephemeris of the 
Solar System bodies. For sake of simplicity we consider here that the 
perturbing bodies move along circular orbits around the barycenter of the 
Sun-Jupiter system; in this case an approximate analytic formula for the 
retarded distance can be given as explained in the following.

From the numerical point of view, an error $\Delta r$ on the reduced distance 
propagates to an error $\Delta h_{00}$ on the metric determination
\[
\Delta h_{00}\lesssim\left|\frac{\mathrm{d}h_{00}}{\mathrm{d}r}\right|
\!\Delta r\simeq h_{00}\frac{\Delta r}{r}.
\]
The model should be accurate to the $(v/c)^3$ order\footnote{%
In this section, where explicit comparisons with non-geometrized quantities 
is required, we will use the $(v/c)^n$ notation for the order of accuracy
instead of the $1/c^n$ one used so far.%
}, so all possible sources of
numerical error must keep this accuracy. We know that $h_{00}\sim(v/c)^2$,
therefore, requiring $\Delta h_{00}<(v/c)^{3}$ implies (see eq.(\ref{eq:metric}))
\[
\frac{\Delta r}{r}<\frac{\tilde{v}}{c}\sim10^{-4}
\]
for the diagonal metric coefficients. Therefore we require that the reduced
distance $r$ to the $a$-th planet be known with a relative error of
\begin{equation}
  \frac{\Delta r}{r}\sim0.1\frac{\tilde{v}_{(a)}}{c}.
  \label{eq:accuracy-conditions}
\end{equation}

From the last consideration it follows that to keep the numerical accuracy
of the tests to the order of $(v/c)^{3}$ inside the Solar System, it
is sufficient to retain only the following terms in the expression of
the retarded distance (\ref{eq:rvector}), i.e.
\begin{equation}
  r=r_{0}+\frac{1}{r_{0}}\left(\mathbf{r}_{0}\cdot\tilde{\mathbf{r}}\right).
  \label{eq:retdist}
\end{equation}
To prove this statement let us start taking planets moving on circular
orbits around the barycenter of the Solar System. Since $\sigma=t+O\left((v/c)^{2}\right)$,
we have
\begin{eqnarray}
  \tilde{x}(t) & = & R\cos(\alpha_{0}+\omega t)\nonumber \\
  \tilde{y}(t) & = & R\sin(\alpha_{0}+\omega t)\label{eq:ephem}\\
  \tilde{z}(t) & = & 0\nonumber
\end{eqnarray}
and consequently
\begin{eqnarray}
  \tilde{x}(t-r) & = & \tilde{x}(t)\cos(\omega r)+\tilde{y}(t)\sin(\omega r)\nonumber \\
  \tilde{y}(t-r) & = & \tilde{y}(t)\cos(\omega r)-\tilde{x}(t)\sin(\omega r)\label{eq:retephem}\\
  \tilde{z}(t-r) & = & 0.\nonumber
\end{eqnarray}

Recalling that
\begin{equation}
  r_{0}\equiv\left(\delta_{ij}\Delta x^{i}(t)\Delta x^{j}(t)\right)^{1/2}=
  \sqrt{\left(x(t)-\tilde{x}(t)\right)^{2}+\left(y(t)-\tilde{y}(t)\right)^{2}+
  \left(z(t)-\tilde{z}(t)\right)^{2}}
  \label{eq:r0}
\end{equation}
\begin{equation}
  \mathbf{r}_{0}\equiv\left\{ x(t)-\tilde{x}(t),y(t)-\tilde{y}(t),z(t)-\tilde{z}(t)\right\}
  \label{eq:r0-vec}
\end{equation}
\begin{equation}
  \tilde{\mathbf{r}}\equiv\left\{\tilde{x}(t)-
  \tilde{x}(t-r),\tilde{y}(t)-\tilde{y}(t-r),\tilde{z}(t)-\tilde{z}(t-r)\right\}
  \label{eq:rtilde-vec}
\end{equation}
and substituting eqs.(\ref{eq:retephem}) to (\ref{eq:rtilde-vec}) into (\ref{eq:retdist})
and after some algebra, we obtain the following expression (in geometrized
units)
\begin{equation}
  r=r_{0}+\frac{1}{r_{0}}\left[\left(x\tilde{x}+y\tilde{y}\right)-
  \left(x\tilde{x}+y\tilde{y}\right)\cos(\omega r)-\left(x\tilde{y}-
  y\tilde{x}\right)\sin(\omega r)-R^{2}\left[1-\cos\left(\omega r\right)\right]\right],
  \label{eq:retdist-circorb}
\end{equation}
where, for the sake of simplicity, we put $x(t)\equiv x$ and $\tilde{x}(t)\equiv\tilde{x}$.

The previous expression can be solved numerically in principle to
any degree of accuracy, but it cannot be solved analytically because
equation (\ref{eq:retdist-circorb}) is transcendent, like, for example,
Kepler's equation.

However, for all the planets in the Solar System, a first-order analytical
expression for $r$ is sufficient to satisfy the requirements set
by (\ref{eq:accuracy-conditions}) about the reduced distance, and
therefore to have the metric coefficients approximated to $(v/c)^{3}$. 
With no loss of generality, we can take a planet whose orbital radius is $R$, 
a series of events $P\equiv(x,y,z,\bar{t})$, and assume that:
\begin{enumerate}
\item the spatial coordinates of $P$ all belong to the $x$-axis (that
is $P=(x,0,0,\bar{t})$);
\item the orbit of the planet is equatorial, i.e. $\tilde{z}(t)=0\;\forall\; t$;
\item the planet has coordinates $\tilde{x}(\bar{t})=R$, $\tilde{y}(\bar{t})=0$.
\end{enumerate}
With these hypothesis $r_{0}=x(\bar{t})-R$, and we expect that $r=r_{0}$
when $r/c\equiv\Delta t=kT$ $(k=1,2,\dots,n)$ where $T$ is the
orbital period of the planet.

\begin{figure}[htbp]
  \plottwo{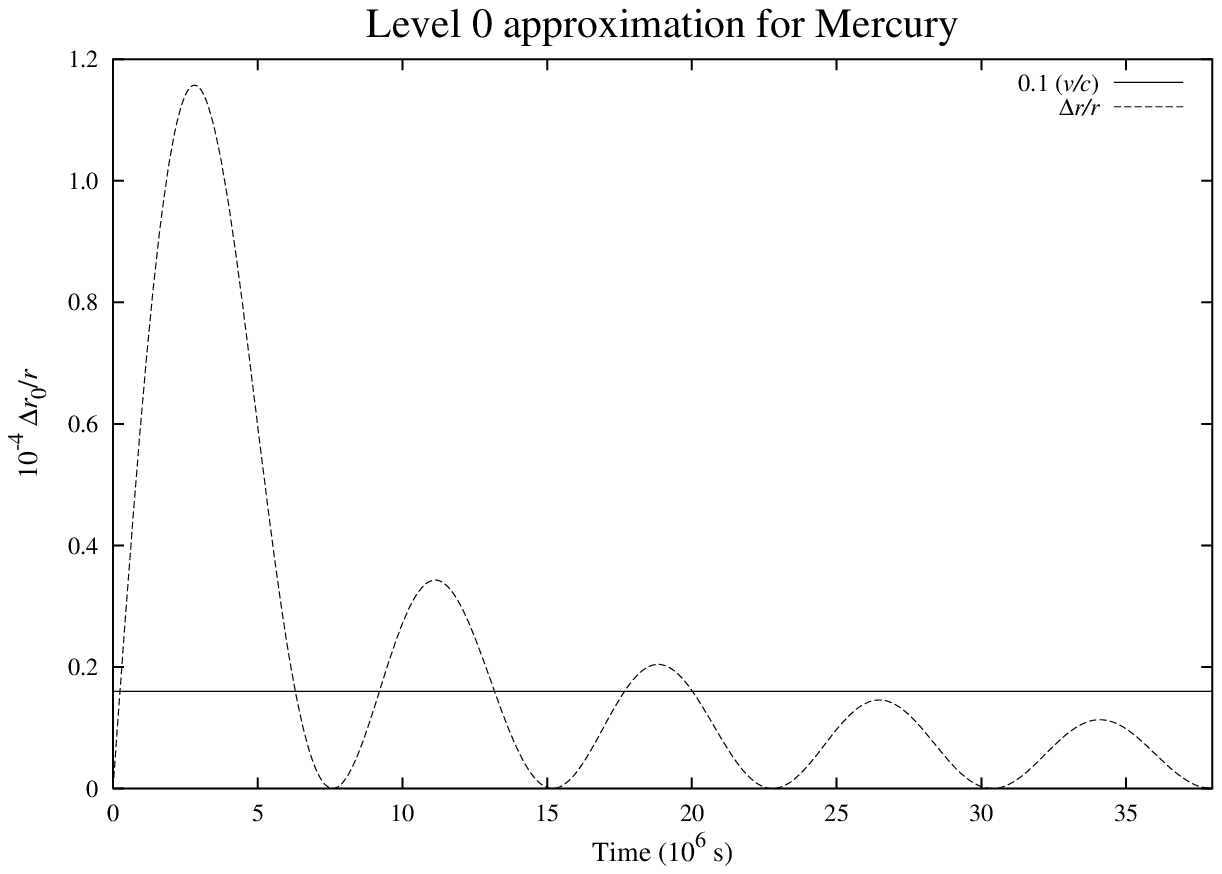}{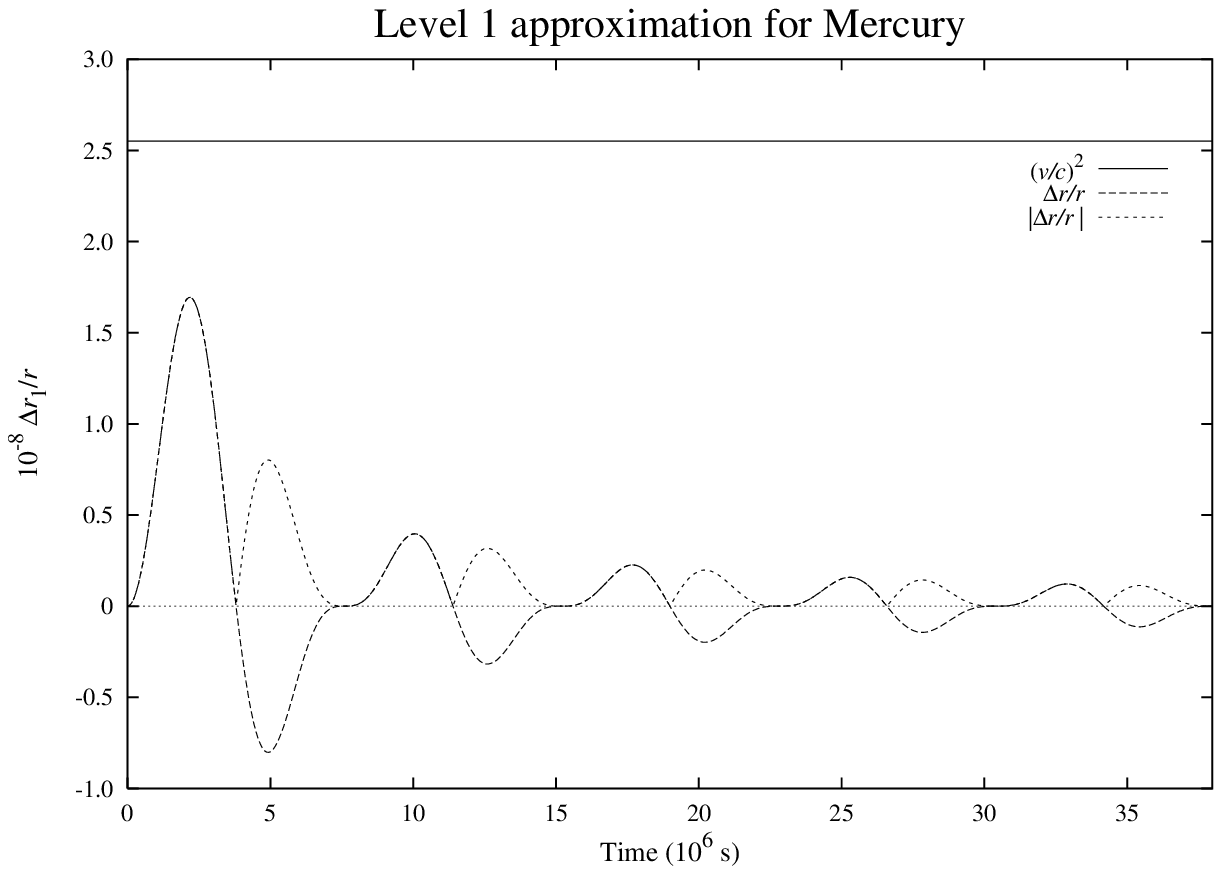}
  \caption{\label{fig:retdist-accuracy}Plot of $\Delta r_{0}/r$ (left) and 
    $\Delta r_{1}/r$ (right). The solid line is 
    $0.1\cdot(v_{\mathrm{mercury}}/c)$, i.e. the floor for $\mathrm{d}r/r$
    according to (\ref{eq:accuracy-conditions}), in the left plot, and 
    $(v_{\mathrm{mercury}}/c)^{2}$ in the right one. The dotted lines indicate 
    the relative errors. The two plots refer to the case of Mercury and are
    representative of all of the other planets, which follow the same trend.
    It can be seen that the zero-order approximation ($r_0$) is not sufficient 
    to get the required numerical accuracy of the reduced distance all along the 
    integration path, while the first order one ($r_1$) keeps the differences
    always well under the $(v/c)^2$ level.}
\end{figure}

Once the series of events $P$ is generated  then putting $x=R+c\Delta t$
and using $\Delta t$ as the independent variable in the range $(0,5T]$
for each planet of the Solar System, we calculate for each $P$:
\begin{enumerate}
\item the {}``exact'' reduced distance $r$ by solving numerically equation
(\ref{eq:rvector});
\item its zero-order approximation, i.e. $r_{0}=(R+c\Delta t)-R$;
\item its first-order approximation obtained from (\ref{eq:retdist-circorb}),
namely
\begin{equation}
  r_{1}=r_{0}+\frac{1}{r_{0}}\left\{ \left(x\tilde{x}+y\tilde{y}\right)-
  \left(x\tilde{x}+y\tilde{y}\right)\cos(\omega r_{0})-
  \left(x\tilde{y}-y\tilde{x}\right)\sin(\omega r_{0})-
  R^{2}\left[1-\cos\left(\omega r_{0}\right)\right]\right\};
  \label{eq:r1}
\end{equation}
\item the relative error for $r_{0}$, $\Delta r_{0}/r=(r-r_{0})/r$;
\item the relative error for $r_{1}$, $\Delta r_{1}/r=(r-r_{1})/r$.
\end{enumerate}
The plots in figure~\ref{fig:retdist-accuracy} show that, as expected, 
$r_{0}$ is not a good approximation for $r$ in the $(v/c)^{3}$ unless $P$ is
at least $\sim3cT$ far from the planet, while $r_{1}$ is a good
approximation since it is always $\Delta r_{1}/r<(v/c)^{2}$.
\clearpage
\clearpage
\subsection{Test design and results}
It can be easily verified that if we put the velocities of the perturbing
bodies to zero, the formulae for RAMOD4 reduce to those of RAMOD3.
Therefore the first thing checked was that the new code produced the
same results as the original RAMOD3 code when $\tilde{\mathbf{v}}^{(a)}=0$.
This cannot be properly called a test, nevertheless it is a basic
sanity check to validate the correct implementation of the model in the
computer code.

After this check, we started the real test phase having in mind that
the proper design of the tests should highlight the effect of the
non-stationarity of the space-time, which marks the main difference
between RAMOD3 and RAMOD4, as thoroughly discussed in the previous
sections.

First of all, the stationary metric of RAMOD3 allowed us to design a
straightforward test of spherical-symmetry, a property peculiar
to such a gravitational field. The same is not possible in RAMOD4;
therefore identical geometrical configurations adapted for RAMOD4 should
produce differences $\lesssim10\;\mu\mathrm{as}$, which is the intrinsic
order of accuracy of RAMOD3.

Moreover, the non-stationariety of the  space-time of RAMOD4 can be conceived
as being due to the contribution of three terms, namely:
\begin{enumerate}
\item the motion of the planets (i.e. the time dependence of the positions
of the bodies $r(t)$);
\item the inclusion of the retarded distance, that takes into account the
finite propagation of gravity (negligible at the order of $1/c^{2}$);
\item the presence of terms to the $1/c^{3}$ order which depends explicitly
on the velocity of the perturbing bodies; these include the off-diagonal
terms of the metric $h_{0i}$ and the retarded corrections proportional to
$\mathbf{v}^{(a)}\cdot\mathbf{n}^{(a)}$ in the diagonal coefficients of the metric.
\end{enumerate}
The above considerations naturally lead us to design two specific
set of tests which are described below. The second one, in particular,
aims at comparing different versions of \emph{codes} from a numerical 
point of view.
\clearpage
\begin{figure}
  \plotone{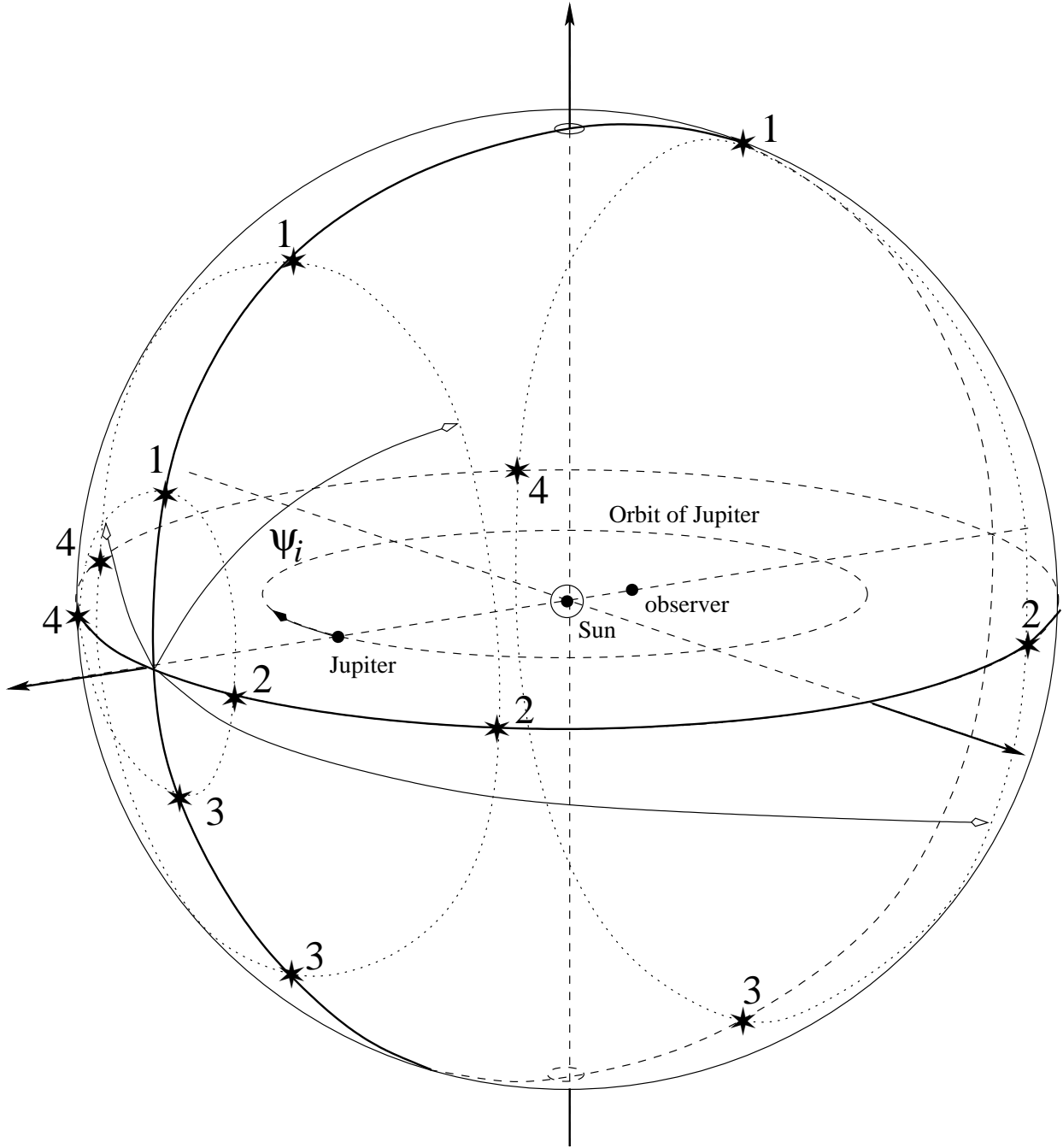}
  \caption{\label{fig:self-cons}Geometry for the self-consistency test. The
    observer is aligned along the Sun-Jupiter direction and for each angular
    distance $\psi$ four stars at symmetric positions with respect to
    the perturbing bodies are considered. In RAMOD4 the spherical symmetry
    is broken by the orbital motion of the bodies. In the figure only the 
    orbit of Jupiter is highlighted, while that of the Sun cannot be drawn
    because of its small radius.}
\end{figure}
\clearpage
\begin{deluxetable}{cccccc}
\tablecaption{Summary of the results for the self-consistency test. The 
  first column shows the angular distance $\psi$ from the Sun (in degrees), 
  the second column is the mass ratio of the planet and the Sun, $\delta\psi$ 
  is the deflection (identical in all the four quadrants) obtained from RAMOD3 
  measured in arcseconds, the $\Delta(\delta\psi_{i})$ are the differences in 
  $\mu\mathrm{as}$ between the deflection of RAMOD4 and RAMOD3 in the $i$-th 
  quadrant, according to the schema depicted in 
  figure~\ref{fig:self-cons}.\label{tab:self-cons}}
\tablehead{
  \colhead{$\psi\;(\mathrm{deg)}$} & \colhead{$m_{J}/m_{\odot}$} & \colhead{$\delta\psi$} &
  \colhead{$\Delta(\delta\psi_{1}),\Delta(\delta\psi_{3})$} & 
  \colhead{$\Delta(\delta\psi_{2})$} & \colhead{$\Delta(\delta\psi_{4})$}}
\startdata
$\simeq0.27(\equiv1\: R_{\odot})$ & $10^{-3}$ & $1.7509921$ & $<10^{-1}$ & $-13.2$ & $13.2$ \\
$"$ & $10^{-4}$ & $1.7507492$ & $"$ & $-1.3$ & $1.3$ \\
$"$ & $10^{-5}$ & $1.7507249$ & $"$ & $-0.1$ & $0.1$ \\
$"$ & $10^{-6}$ & $1.7507224$ & $"$ & $<10^{-1}$ & $<10^{-1}$ \\
$1$ & $10^{-3}$ & $0.4667362$ & $"$ & $-1.0$ & $1.0$ \\
$"$ & $10^{-4}$ & $0.4666715$ & $"$ & $-0.1$ & $0.1$ \\
$"$ & $10^{-5}$ & $0.4666651$ & $"$ & $<10^{-1}$ & $<10^{-1}$ \\
$2$ & $10^{-3}$ & $0.2333501$ & $"$ & $-0.2$ & $0.2$ \\
$"$ & $10^{-4}$ & $0.2333177$ & $"$ & $<10^{-1}$ & $<10^{-1}$ \\
$5$ & $10^{-3}$ & $0.0932902$ & $"$ & $<10^{-1}$ & $<10^{-1}$ \\
\enddata
\end{deluxetable}
\clearpage
\paragraph{Self-consistency test.}
Let us remind the original configuration of the spherical symmetry
test done in RAMOD3, i.e. the self-consistency test \citep{2004ApJ...607..580D}.
In the cited paper we imposed the condition that the Sun were the
only source of gravity and looked for the symmetry of the light deflections
with respect to its center. Here we considered the similar case where,
however, the perturbing field is that of the point-like masses of
the Sun and Jupiter aligned with the observer; the geometrical configuration
relative to this test is sketched in figure~\ref{fig:self-cons}.
Here we considered an observer at 1~AU from the Sun and 6.2~AU from
Jupiter, and a set of stars placed at different angular distances $\psi$
from the Sun. For each $\psi$ we have taken four stars symmetrical
positioned with respect to the instantaneous axis joining the observer
to the Sun and Jupiter: the anisotropy among the deflections are 
expected to be more evident on the orbital plane of the two bodies.
The results are reported in table~\ref{tab:self-cons}, where we
can see that both our predictions were satisfied, i.e.:
\begin{enumerate}
\item $\delta\psi_{1}$ and $\delta\psi_{3}$ are equal and coincident 
with those of RAMOD3 to $0.1\,\mu$as, while $\delta\psi_{2}\neq\delta\psi_{4}$ 
and different from those of RAMOD3;
\item the differences between $\delta\psi_{2}$ and $\delta\psi_{4}$ in
RAMOD3 and RAMOD4 are of the expected order ($\simeq13.2\:\mu\mathrm{as}$
for Sun-grazing rays and quickly falling below the $0.1\:\mu\mathrm{as}$
for angular distances $\psi\gtrsim 5$~deg);
\item the deflection is greater in the case where the Sun trajectory is
approaching the photon path and smaller in the opposite side, where
it is getting farther from it.
\end{enumerate}
As a final verification for this self-consistency test, we repeated
several times the same run but with a less massive planet. Obviously,
we expected that, as the mass of the planet decreases, the results
tend to those of the perfectly symmetric case: {\em i.e.} as 
$m_{\mathrm{jup}}\rightarrow0$, the velocities of Jupiter and the Sun 
go to zero as well. Table~\ref{tab:self-cons} shows that the 
deflections are perfectly symmetric down to the $0.1\:\mu\mathrm{as}$ 
level when $m_{\mathrm{planet}}=10^{-3}m_{\mathrm{jup}}$.
\clearpage
\begin{table}
  \begin{center}
   \begin{tabular}{|c|ccc|}
      \cline{1-4} 
      Codes & \multicolumn{3}{|c|}{Contributions} \\\cline{2-4}
      & $r(t)$ & $r_{\mathrm{ret}}$ & $v$ \\\tableline
      \multicolumn{1}{|c|}{R3a} & Yes & No & No \\\cline{1-4}
      \multicolumn{1}{|c|}{R3b} & Yes & Yes & No \\\cline{1-4} 
      \multicolumn{1}{|c|}{R4\phantom{b}} & Yes & Yes & Yes \\\tableline
    \end{tabular}
  \end{center}
  \caption{\label{tab:models}Schematic representation of the different codes
    under comparisons. While R4 stands for the one that implements RAMOD4,
    R3a and R3b indicate different ``flavors'' of a RAMOD3-like code.
    Among the specific contributions, $r(t)$ means that we consider the
    equations of motion of RAMOD3, but taking into account the planetary
    motion in the sense described in the main text of the article; $r_{\mathrm{ret}}$
    that, when calculating the distances (and the velocities) of the perturbing
    bodies from the photon which enter the metric coefficients, we consider
    the finite gravity speed (i.e. the reduced distance); finally, $v$
    means that the equations of motion are those of RAMOD4 and that the
    metric is complete with its $1/c^{3}$ terms depending explicitly on the velocity
    of the perturbing bodies. In the original RAMOD3 code, which strictly implements 
    the $1/c^{2}$ model and its hypotheses, none of these contributions are taken 
    into account, while all of them are present in R4 at its full extent.}
\end{table}
\clearpage
\paragraph{Disentangling the non-stationary space-time effects.}
To this purpose we have checked RAMOD4 against different (somewhat
hybrid) versions of the RAMOD framework, where, according to table~\ref{tab:models},
they will be called in the following:
\begin{enumerate}
\item R4: the code implementing the full RAMOD4 model;
\item R3b: the R4 code without the $1/c^{3}$ terms depending explicitly on 
the velocity of the perturbing bodies. In this case the master equations are 
reconducted to that of RAMOD3, whereas the reduced distances are still fully 
considered and the  planets are moving along their orbits;
\item R3a: the R4 code without both the velocity-dependent terms and the reduced 
distances. In this case we consider, for each step of integration, a different
position of each perturbing body. These positions are those for circular
orbits at time $t_{i}$ of the $i$-th step of integration.
\end{enumerate}
The importance of these tests, as said above, is numerical rather
than physical since, e.g., the exclusion of the reduced distance
cannot be acceptable in a strictly physical sense. It can be important,
however, if one is interested in the practical implementation of a
particular code: in this case what one should care about
is the efficiency and the speed of the code given the required level
of accuracy. A similar comparison was done by \citet{2003A&A...410.1063K}
for the model in \citet{2003AJ....125.1580K}, and in this sense the
R3a and R3b cases resembles cases $P_{1}$ and $P'_{3}$ of that paper,
respectively.

\begin{figure}
  \plotone{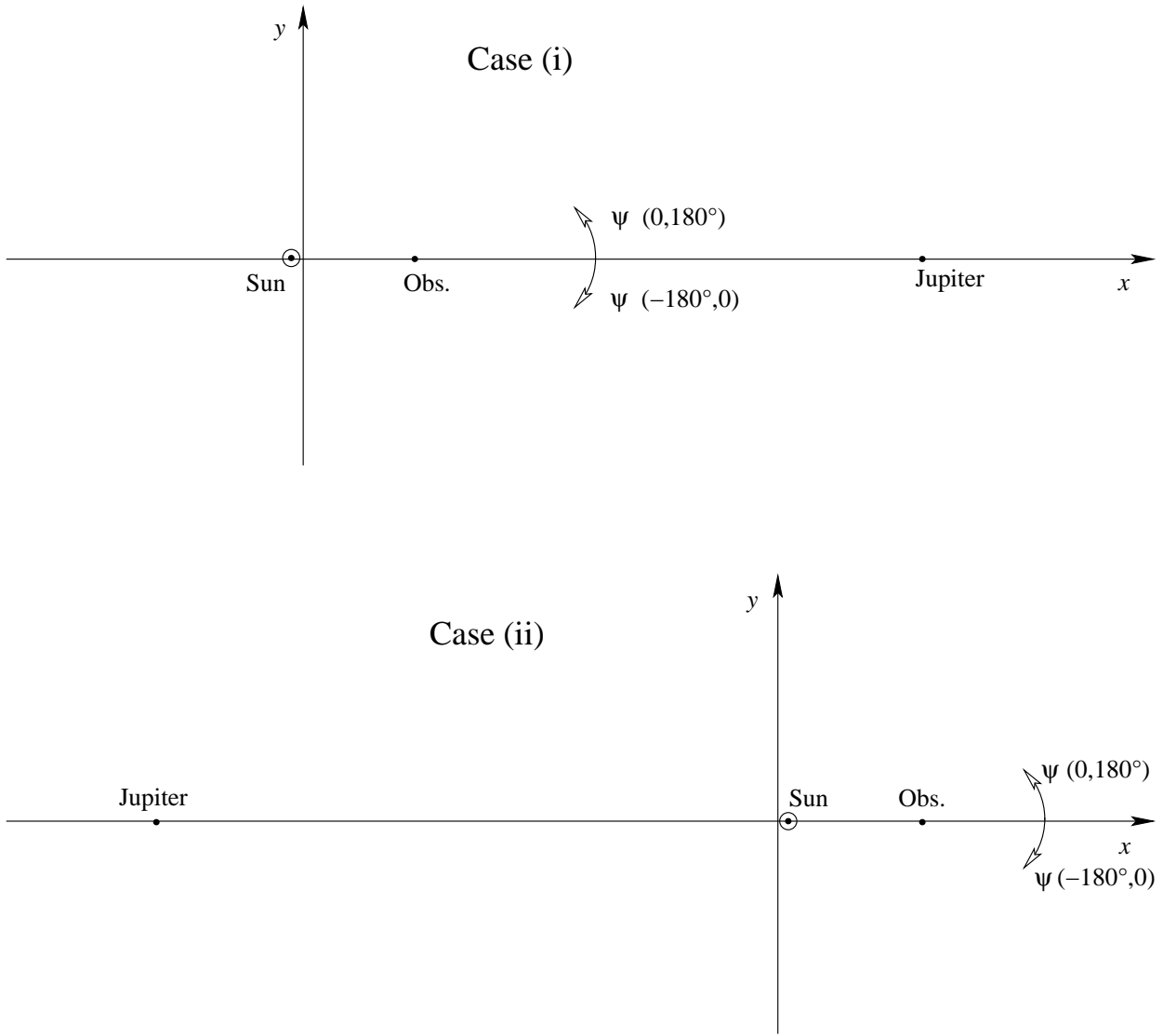}
  \caption{\label{fig:geometry}Sketch of the geometrical configurations for
    the tests using the codes R4, R3b and R3a.}
\end{figure}

In these tests we considered only the presence of the Sun and Jupiter,
since all the planets can be added with no loss of generality in the
same way as Jupiter. Two cases have been taken into account, as sketched
in figure~\ref{fig:geometry}: (i) when the observer lies between
the Sun and Jupiter, or (ii) when the Sun is between the observer and
Jupiter. The first case allows to evaluate the specific contribution
of the Sun and of Jupiter separately, putting them in opposite direction
with respect to the observer, while the second is more sensitive in
order to investigate at which level of accuracy the model feels the
non-linearity of the superimposition of the two-body gravitational
field.
\clearpage
\clearpage
For both of the cases the results indicate that:
\begin{enumerate}
\item the differences between R3a and R4 are $\sim4.5\:\mu\mathrm{as}$ at most,
meaning that the influence of the fixed planetary positions are of
$\lesssim9\:\mu\mathrm{as}$. In particular, the greatest effect is for
photons which graze Jupiter ($\sim4.5\:\mu\mathrm{as}$) instead of
the Sun ($\lesssim0.3\:\mu\mathrm{as}$);
\item the differences between R3b and R4 are tipically less than those 
between R3a and R4, but their ratios vary according to the geometrical 
configuration of the perturbing bodies and the observer, going from, e.g., one
order of magnitude to about $1$. Once again the greatest difference occurs in 
the case of photons grazing Jupiter, as it should be expected since its velocity 
is far bigger than that of the Sun, and it amounts to $\sim1.8\:\mu\mathrm{as}$ 
while the Sun contributes with $\sim10^{-1}\:\mu\mathrm{as}$ at most.
\end{enumerate}
This means that one should be careful at choosing the code integrating the
photons' geodesics. Optimizations are possible, but they must be tuned with
the geometrical configuration of the perturbing bodies.
\clearpage
\begin{deluxetable}{rcccrr}
\tabletypesize{\scriptsize}
\tablecaption{Excerpts from the results of the code-comparing
test. This table is for the (i) case, in which the observer lies
between the Sun and Jupiter. The first column gives the angle that the 
photon's incoming direction forms with the $x$ axis counted counter-clockwise in the
range $[-180^\circ,180^\circ)$ as sketched in figure~\ref{fig:geometry}.
The next three columns contain the deflections given by the three
codes (in arcseconds) and the fifth and sixth columns give the difference
in $\mu\mathrm{as}$ of R3b and R3a with R4 respectively.\label{tab:disent-i}}
\tablehead{
  \colhead{$\psi\;(\mathrm{deg)}$} & \colhead{} &
  \colhead{$\delta\psi\;\mathrm{(arcsec)}$} & \colhead{} & 
  \colhead{$(\delta\psi_{\mathrm{R4}}-\delta\psi_{\mathrm{R3b}})\;(\mu\mathrm{as})$} &
  \colhead{$(\delta\psi_{\mathrm{R4}}-\delta\psi_{\mathrm{R3a}})\;(\mu\mathrm{as})$}}
\startdata
 & R4 & R3b & R3a & & \\\tableline
179.7334367 & 1.75073792543 & 1.75073781174 & 1.75073764027 & 0.11369 & 0.28516 \\
179.4668735 & 0.87535666435 & 0.87535660782 & 0.87535652209 & 0.05653 & 0.14226 \\
178.4006205 & 0.29176702035 & 0.29176700193 & 0.29176697336 & 0.01842 & 0.04699 \\
160.1792868 & 0.02330937402 & 0.02330937317 & 0.02330937094 & 0.00086 & 0.00308 \\
120.1507651 & 0.00707469758 & 0.00707469781 & 0.00707469723 & -0.00024 & 0.00035 \\
80.1222434 & 0.00342350034 & 0.00342350079 & 0.00342350051 & -0.00045 & -0.00018 \\
40.0937217 & 0.00148347654 & 0.00148347699 & 0.00148347671 & -0.00044 & -0.00016 \\
5.7835603 & 0.00018740032 & 0.00018739969 & 0.00018739782 & 0.00062 & 0.00250 \\
0.03912 & 0.0025469505 & 0.0025471242 & 0.00254738571 & -0.17370 & -0.43521 \\
0.01956 & 0.00480767427 & 0.00480800257 & 0.00480849601 & -0.32830 & -0.82174 \\
0.00652 & 0.01176284745 & 0.01176365151 & 0.01176485859 & -0.80406 & -2.01115 \\
-0.00652 & 0.0263647168 & 0.02636291438 & 0.02636020825 & 1.80242 & 4.50856 \\
-0.01956 & 0.0062147205 & 0.00621429593 & 0.00621365819 & 0.42457 & 1.06231 \\
-0.03912 & 0.00289458627 & 0.00289438879 & 0.00289409164 & 0.19749 & 0.49463 \\
-5.7835603 & 0.00018738265 & 0.00018738327 & 0.00018738515 & -0.00062 & -0.00250 \\
-40.0937217 & 0.00148347648 & 0.00148347603 & 0.00148347631 & 0.00044 & 0.00016 \\
-80.1222434 & 0.00342350043 & 0.00342349998 & 0.00342350025 & 0.00045 & 0.00018 \\
-120.1507651 & 0.0070746969 & 0.00707469666 & 0.00707469725 & 0.00024 & -0.00035 \\
-160.1792868 & 0.02330936614 & 0.02330936699 & 0.02330936922 & -0.00085 & -0.00308 \\
-178.4006205 & 0.29176611396 & 0.29176613238 & 0.29176616095 & -0.01842 & -0.04699 \\
-179.4668735 & 0.87534872947 & 0.87534878599 & 0.87534887173 & -0.05653 & -0.14226 \\
-179.7334367 & 1.75070641203 & 1.75070652571 & 1.75070669718 & -0.11368 & -0.28515 \\
\enddata        
\end{deluxetable}
                
\begin{deluxetable}{rcccrr}
\tabletypesize{\scriptsize}
\tablecaption{Excerpts from the results of the code-comparing
test. This table is for the (ii) case, in which the Sun is between the observer 
and Jupiter. The table headings are the same as in table~\ref{tab:disent-i}.\label{tab:disent-ii}}
\tablehead{     
  \colhead{$\psi\;(\mathrm{deg)}$} & \colhead{} &
  \colhead{$\delta\psi\;\mathrm{(arcsec)}$} & \colhead{} & 
  \colhead{$(\delta\psi_{\mathrm{R4}}-\delta\psi_{\mathrm{R3b}})\;(\mu\mathrm{as})$} &
  \colhead{$(\delta\psi_{\mathrm{R4}}-\delta\psi_{\mathrm{R3a}})\;(\mu\mathrm{as})$}}
\startdata      
 & R4 & R3b & R3a & & \\\tableline
179.7334367 & 1.75097892390 & 1.75097901914 & 1.75097916265 & -0.09525 & -0.23876 \\
179.4668735 & 0.87548418970 & 0.87548423717 & 0.87548430900 & -0.04746 & -0.11930 \\
178.4006205 & 0.29181111968 & 0.29181113523 & 0.29181115918 & -0.01554 & -0.03950 \\
160.1792868 & 0.02331311678 & 0.02331311760 & 0.02331311946 & -0.00082 & -0.00268 \\
120.1507651 & 0.00707631845 & 0.00707631834 & 0.00707631882 & 0.00011 & -0.00036 \\
80.1222434 & 0.00342512778 & 0.00342512747 & 0.00342512763 & 0.00032 & 0.00016 \\
40.0937217 & 0.00148624154 & 0.00148624114 & 0.00148624118 & 0.00040 & 0.00037 \\
5.7835603 & 0.00020574544 & 0.00020574502 & 0.00020574502 & 0.00043 & 0.00042 \\
0.0391200 & 0.00000139069 & 0.00000139027 & 0.00000139027 & 0.00043 & 0.00043 \\
-0.0391200 & 0.00000139027 & 0.00000139069 & 0.00000139069 & -0.00043 & -0.00043 \\
-5.7835603 & 0.00020574502 & 0.00020574544 & 0.00020574544 & -0.00043 & -0.00042 \\
-40.0937217 & 0.00148624118 & 0.00148624158 & 0.00148624155 & -0.00040 & -0.00037 \\
-80.1222434 & 0.00342512767 & 0.00342512799 & 0.00342512783 & -0.00032 & -0.00016 \\
-120.1507651 & 0.00707631909 & 0.00707631920 & 0.00707631873 & -0.00011 & 0.00036 \\
-160.1792868 & 0.02331312349 & 0.02331312268 & 0.02331312081 & 0.00082 & 0.00268 \\
-178.4006205 & 0.29181187997 & 0.29181186441 & 0.29181184044 & 0.01555 & 0.03952 \\
-179.4668735 & 0.87549084388 & 0.87549079633 & 0.87549072436 & 0.04755 & 0.11952 \\
-179.7334367 & 1.75100533870 & 1.75100524311 & 1.75100509907 & 0.09560 & 0.23963 \\
\enddata
\end{deluxetable}
\clearpage
\section{Conclusions}
Aim of modern relativistic astrometry is to produce a three dimensional
rendering of our Galaxy with an accuracy of one $\mu\mathrm{as}$ in the
measurements of angles. In order to reach this goal one needs an appropriate
algorithm to trace back to their emission space-time points the light
signals which are intercepted by a suitable observational device.

In the near future, one of such devices will be the astrometric satellite 
Gaia which will orbit the Sun from nearby the Sun-Earth outer Lagrangian point $L_{2}$.
The expected accuracy of the satellite observations requires that one 
should treat the light propagation through
the Solar System in a general relativistic framework and consider
the general relativistic effects induced by the bodies of the Solar
System up to the order of $1/c^{3}$.

Our strategy was to construct a series of models with increasing generality
and complexity, exploiting each of them as a test-bed for the more
advanced one. Our last model accurate to $1/c^{3}$ is RAMOD4. As
previously discussed, we first compared RAMOD4 to its less accurate predecessor
RAMOD3, which was accurate to $1/c^{2}$. With a suitable handling of
the items considered in the model, we are able to highlight the relative
importance of the individual sources of relativistic perturbations
and judge about their physical relevance.

Our first conclusion is that RAMOD4 behaves as expected with respect to RAMOD3,
that is, the differences between the two models are at the expected level of 
$\sim10~\mu\mathrm{as}$. So this work successfully completes the series of
models from a theoretical point of view and represents a launching platform to
deduce with a completely numerical algorithm the astrometric parameters of a
celestial object from a well-defined set of relativistically measured
quantities.

This model, however, is part of a broader project whose completion requires its 
implementation into a feasible and efficient structure for the data reduction
of the Gaia mission. In this sense, the practical application of this model
requires also to investigate the specific ways of implementing it with respect
to the accuracy goals of the mission. For this reason, then, we tried to single
out the different contribution which come out together from the motion of the
perturbing bodies, and evaluate them separately. We found that the contribution 
to the total deflection coming from the velocity-induced terms of the metric is 
generally smaller than that of the reduced distance, but not negligible at the 
$0.1~\mu\mathrm{as}$ level. The former one can in fact reach a maximum amount of 
$\sim1.8\:\mu\mathrm{as}$ while the latter goes up to $\sim4.5\:\mu\mathrm{as}$.
Both these cases, and the greatest differences with RAMOD3, come out when the 
photon is grazing Jupiter, while the Sun contributes up to less than 
$0.3~\mu\mathrm{as}$ that is one order of magnitude lower.


Given the unprecedented accuracy reachable by Gaia-like missions, several new
relativistic tests related to the light propagation in a non-stationary
gravitational field could be carried out. From this point of view it is
advisable to create a cross-check framework of models to validate the
relativistic outputs induced by the modeling itself at the microarcsecond level. 
In fact, our Relativistic Astrometric Model naturally confronts itself with the
works of Kopeikin, Sch\"afer and Mashhoon \citep{1999PhRvD..60f124002K,2002PhRvD..65f4025K} 
and \citet{2003AJ....125.1580K}. While the latter stems from the general
formulation of the former conforming to its basic tenets, our model
responds to an inherently different strategy. As far as the astrometric
problem is concerned, in the Kopeikin-Sch\"afer and Klioner approach the light
ray is reconstructed as a sum of terms which allows for a direct evaluation of
the individual relativistic effects induced by the gravitating bodies that the
light ray is expected to encounter on its way. Each relativistic effect enters
as a perturbation of the light ray trajectory and is treated to a sufficiently
high order (not less than $1/c^3$) to cope with the accuracy expected from
Gaia's observations.  Our model instead aims to determine a full solution for
the light trajectory which naturally includes, in a curved space-time, all the 
individual effects; the latters are somewhat hidden in the covariant formalism
of our approach and directly contribute to the solutions of our master
equations.  Evidently any specific effect one is interested to explore can be
independently deduced from our formalism as a branch output, once we adapt the
master equation to a required specific case.

Although our model and Klioner's are of comparable accuracy as it can be 
deduced from individual tests, the overall structure of Klioner's model makes 
it ready to handle the reduction of a large number of observational data in a 
computational efficient way; nevertheless we are confident that our model will 
soon reach the same versatility and operate as an essential tool of verification. 
However, the general covariant formulation of our model, including the complete 
relativistic treatment of the satellite attitude 
(\citealp{2003CQGra..20.2251B,2003CQGra..20.4695B}), represents a well defined 
framework where any desired advancement in the light tracing problem using 
astrometric data is contemplated.

\acknowledgements
This work has been partially supported by the Italian Space Agency
(ASI) under contracts ASI I/R/117/01, and by the Italian Ministry
for Research (MIUR) through the COFIN 2001 program. Vecchiato acknowledges the 
support of the National Institute for Astrophysics (INAF) and Crosta the 
Astronomical Observatory of Torino (INAF-OATo).

\appendix
\section{The attitude frame\label{app:lb}}
Given the tetrad adapted to the local barycentric observer $\{\boldsymbol{u},\boldsymbol{\lambda}_{\hat{a}}\}$
\citep{2003CQGra..20.2251B}, first we identify the spatial direction
to the Sun as seen in the local BCRS and at a general point on the
satellite's space-time trajectory. Then we construct the triad $\{\underset{\mathrm{s}}{\boldsymbol{\lambda_{\hat{a}}}}\}$
where one of those vectors identifies the Sun direction. Clearly the
set $\{\boldsymbol{u},\underset{\mathrm{s}}{\boldsymbol{\lambda}_{\hat{a}}}\}$
forms still an orthonormal tetrad adapted to the local barycentric
observer.

As second step, we boost the vectors of the triad $\{\underset{\mathrm{s}}{\boldsymbol{\lambda}_{\hat{a}}}\}$
to the satellite rest frame; remembering that $\boldsymbol{u}'$ is
the vector field tangent to Gaia's world-line we have \citep{1992AnnPhys.215..1J}:

\begin{equation}
  \underset{\mathrm{bs}}{\lambda_{\hat{a}}^{\alpha}}=
    P(u')^{\alpha}{}_{\sigma}\left[\underset{\mathrm{s}}
    {\lambda_{\hat{a}}^{\sigma}}-\frac{\gamma}{\gamma+1}
    \nu^{\sigma}\left(\nu^{\rho}\underset{\mathrm{s}}
    {\lambda_{\rho\hat{a}}}\right)\right]_{{\hat{a}}=1,2,3}.
  \label{eq:boost}
\end{equation}

The tetrad $\{\underset{\mathrm{bs}}{\boldsymbol{\lambda}_{\hat{0}}}\equiv\boldsymbol{u}',
\underset{\mathrm{bs}}{\boldsymbol{\lambda}_{\hat{a}}}\}$
represents a CoMRS \textit{Sun-locked frame}, \textit{i.e.} a Center-of-Mass
Reference System (\citealp{2004rcn.tech.ll003B}) comoving with the
satellite and with one axis fixed toward the Sun at any point of its
Lissajous orbit around $L_{2}$. The relation between the components
$\nu^{\alpha}$ of the spatial four-velocity $\boldsymbol{\nu}$ and
the components $\beta^{i}$ appearing in (\ref{eq:us}) is easily
established from (\ref{eq:us}) itself and (\ref{eq:ni}) and reads:

\begin{equation}
  \nu^{\alpha}=\frac{1}{\gamma}\left[u'^{0}\left(\beta_{i}\delta^{i\alpha}+
  \delta^{0\alpha}\right)-u^{\alpha}\gamma\right].
\end{equation}
The explicit expressions of the components of these vectors relative
to the BCRS are reported in \citet{2003CQGra..20.4695B}.

 The triad components are given by
\begin{eqnarray}
  \underset{\mathrm{bs}}{\lambda_{\hat{1}}^{t}} & = & \Psi(1+\frac{1}{2}\beta^{2}+3U)+
    \cos\theta\cos\phi h_{01}+\cos\theta\sin\phi h_{02}+\sin\theta h_{03}\label{eq:Lbs1t}\\
  \underset{\mathrm{bs}}{\lambda_{\hat{1}}^{x}} & = & \cos\theta\cos\phi(1-U)+
    \frac{1}{2}\beta_{1}\Psi\label{eq:Lbs1x}\\
  \underset{\mathrm{bs}}{\lambda_{\hat{1}}^{y}} & = & \cos\theta\sin\phi(1-U)+
    \frac{1}{2}\beta_{2}\Psi\label{eq:Lbs1y}\\
  \underset{\mathrm{bs}}{\lambda_{\hat{1}}^{z}} & = & \sin\theta(1-U)+
    \frac{1}{2}\beta_{3}\Psi\label{eq:Lbs1z}\\
  \underset{\mathrm{bs}}{\lambda_{\hat{2}}^{t}} & = & \Phi(1+\frac{1}{2}\beta^{2}+3U)-
    \sin\phi h_{01}+\cos\phi h_{02}\label{eq:Lbs2t}\\
  \underset{\mathrm{bs}}{\lambda_{\hat{2}}^{x}} & = & -\sin\phi(1-U)+
    \frac{1}{2}\beta_{1}\Phi\label{eq:Lbs2x}\\
  \underset{\mathrm{bs}}{\lambda_{\hat{2}}^{y}} & = & \cos\phi(1-U)+
    \frac{1}{2}\beta_{2}\Phi\label{eq:Lbs2y}\\
  \underset{\mathrm{bs}}{\lambda_{\hat{2}}^{z}} & = & -\frac{1}{2}\beta_{3}\Phi\label{eq:Lbs2z}\\
  \underset{\mathrm{bs}}{\lambda_{\hat{3}}^{t}} & = & \Theta(1+\frac{1}{2}\beta^{2}+3U)-
    \sin\theta(\cos\phi h_{01}+\sin\phi h_{02})+\cos\theta h_{03}\label{eq:Lbs3t}\\
  \underset{\mathrm{bs}}{\lambda_{\hat{3}}^{x}} & = & -\cos\theta\sin\theta(1-U)+
    \frac{1}{2}\beta_{1}\Theta\label{eq:Lbs3x}\\
  \underset{\mathrm{bs}}{\lambda_{\hat{3}}^{y}} & = & -\sin\theta\sin\phi(1-U)+
    \frac{1}{2}\beta_{2}\Theta\label{eq:Lbs3y}\\
  \underset{\mathrm{bs}}{\lambda_{\hat{3}}^{z}} & = & \cos\theta(1-U)+
    \frac{1}{2}\beta_{3}\Theta\label{eq:Lbs3z}
\end{eqnarray}
where
\begin{eqnarray}
  \Psi & \equiv & \cos\theta\cos\phi\beta_{1}+\cos\theta\sin\theta\beta_{2}+\sin\theta\beta_{3}\\
  \Phi & \equiv & -\sin\phi\beta_{1}+\cos\phi\beta_{2}\\
  \Theta & \equiv & -\sin\theta(\cos\phi\beta_{1}+\sin\phi\beta_{2})+\cos\theta\beta_{3}.
\end{eqnarray}
Here $2U=h_{00}$ and the angles $\theta$ and $\phi$ are defined as
\begin{equation}
  \phi=\tan^{-1}\frac{y'_{\odot}}{x'_{\odot}},\qquad\theta=
  \tan^{-1}\frac{z'_{\odot}}{\sqrt{x'{}_{\odot}^{2}+y'{}_{\odot}^{2}}}
\end{equation}
where $x'_{\odot}$, $y'_{\odot}$ and $z'_{\odot}$ fix the spatial
coordinate position of the Sun relative to the satellite at each time
$t$; clearly in terms of the coordinates of the Sun and the satellite
they are defined as
\begin{equation}
  x'_{\odot}=x_{\odot}-x_{0}\qquad y'_{\odot}=y_{\odot}-y_{0}\qquad z'_{\odot}=z_{\odot}-z_{0}.
  \label{eq:sun-coord}
\end{equation}
See \citet{2003CQGra..20.4695B} for details. It is straightforward
although tedious to show analytically that $\bar{\ell}_{0}^{k}$ as
solution of (\ref{eq:ea}) with triad (\ref{eq:E1}), (\ref{eq:E2}),
(\ref{eq:E3}) is unitary as expected.

\section{Explicit form for the master equations\label{app:me}}
Adopting the metric given in (\ref{eq:h-approx}), the differential equations for the spatial components
$\bar{\ell}^{i}$ of the null geodesic take the following form:\arraycolsep=0pt
\begin{eqnarray*}
\frac{\mathrm{d}^{2}x^{k}}{\mathrm{d}t^{2}}+\sum_{a} & \left\{\displaystyle{\frac{2m^{(a)}}{{r^{(a)}}^{2}}}\right. & \left[\frac{1}{2}\left(\frac{\mathrm{d}x^{k}}{\mathrm{d}t}\frac{\partial r^{(a)}}{\partial t}+\frac{\partial r^{(a)}}{\partial x^{k}}\left(1+\mathbf{v}^{(a)}\cdot\mathbf{n}^{(a)}\right)\right)\left(\left(\frac{\mathrm{d}x}{\mathrm{d}t}\right)^{2}+\left(\frac{\mathrm{d}y}{\mathrm{d}t}\right)^{2}+\left(\frac{\mathrm{d}z}{\mathrm{d}t}\right)^{2}\right)\right.\\
 &  & -\left(\frac{3}{2}\frac{\mathrm{d}x^{k}}{\mathrm{d}t}\left(1+\mathbf{v}^{(a)}\cdot\mathbf{n}^{(a)}\right)-2v^{(a)}_{k}\right)\left(\frac{\mathrm{d}x}{\mathrm{d}t}\frac{\partial r^{(a)}}{\partial x}+\frac{\mathrm{d}y}{\mathrm{d}t}\frac{\partial r^{(a)}}{\partial y}+\frac{\mathrm{d}z}{\mathrm{d}t}\frac{\partial r^{(a)}}{\partial z}\right)\\
 &  & \left.-\frac{\mathrm{d}x^{k}}{\mathrm{d}t}\frac{\partial r^{(a)}}{\partial t}+\frac{1}{2}\frac{\partial r^{(a)}}{\partial x^{k}}\left(1+\mathbf{v}^{(a)}\cdot\mathbf{n}^{(a)}\right)-2\frac{\partial r^{(a)}}{\partial x^{k}}\left(\frac{\mathrm{d}x}{\mathrm{d}t}v^{(a)}_{x}+\frac{\mathrm{d}y}{\mathrm{d}t}v^{(a)}_{y}+\frac{\mathrm{d}z}{\mathrm{d}t}v^{(a)}_{z}\right)\right]+\\
 & \displaystyle{\frac{2m^{(a)}}{r^{(a)}}} & \left[\frac{3}{2}\frac{\mathrm{d}x^{k}}{\mathrm{d}t}\left(\frac{\mathrm{d}x}{\mathrm{d}t}\frac{\partial \left(\mathbf{v}^{(a)}\cdot\mathbf{n}^{(a)}\right)}{\partial x}+\frac{\mathrm{d}y}{\mathrm{d}t}\frac{\partial \left(\mathbf{v}^{(a)}\cdot\mathbf{n}^{(a)}\right)}{\partial y}+\frac{\mathrm{d}z}{\mathrm{d}t}\frac{\partial \left(\mathbf{v}^{(a)}\cdot\mathbf{n}^{(a)}\right)}{\partial z}\right)-\right.\\
  &  & \left.\frac{1}{2}\frac{\partial\left(\mathbf{v}^{(a)}\cdot\mathbf{n}^{(a)}\right)}{\partial x^{k}}\left(\left(\frac{\mathrm{d}x}{\mathrm{d}t}\right)^{2}+\left(\frac{\mathrm{d}y}{\mathrm{d}t}\right)^{2}+\left(\frac{\mathrm{d}z}{\mathrm{d}t}\right)^{2}+1\right)\right]-\\
 & \displaystyle{\frac{4m^{(a)}}{r^{(a)}}} & \left.\frac{\mathrm{d}\mathbf{r}}{\mathrm{d}t}\cdot\left(\nabla\times\mathbf{v}^{(a)}\right)_{k}\right\}=0\\
\end{eqnarray*}

In the case of circular orbits, the last line of each equation disappears since $\nabla\times\mathbf{v}^{(a)}=0$ in this case. Moreover with straightforward calculations, similar to those appeared in section~\ref{sub:implementing-reduced}, one can easily find that
\[
\mathbf{v}^{(a)}\cdot\mathbf{n}^{(a)}=\frac{\omega}{r^{(a)}}\left[(yx^{(a)}-xy^{(a)})\cos(\omega r^{(a)})+(xx^{(a)}+yy^{(a)})\sin(\omega r^{(a)})\right]
\]
and
\begin{eqnarray}
\frac{\partial\left(\mathbf{v}^{(a)}\cdot\mathbf{n}^{(a)}\right)}{\partial x^{k}}&=&\frac \omega{r^{(a)}}
\left[\cos(\omega r^{(a)})(\delta^y_i x^{(a)}-\delta^x_i y^{(a)})+\sin(\omega r^{(a)})(\delta^x_i x^{(a)}+
\delta^y_i y^{(a)})\right]+\nonumber\\
&+&(\partial_ir^{(a)})\left\{\frac{\omega^2}{r^{(a)}}\left[-(y x^{(a)}-x y^{(a)})\sin(\omega r^{(a)})+
(x x^{(a)}+y y^{(a)})\cos(\omega r^{(a)})\right]-
\frac{\mathbf{v}^{(a)}\cdot {\mathbf {r}^{(a)}}}{r^{(a)2}}\right\}\nonumber.
\end{eqnarray}
where all the coordinates of the planets are at time $t$, while $r^{(a)}$ and $\mathbf{v}^{(a)}$ are at 
time $t'$ (this means that one should replace $r^{(a)}$ with its approximation of eq.(\ref{eq:r1}).


\begin{thebibliography}{}

\bibitem[{Bastian}, 2004]{2004rcn.tech.ll003B}
{Bastian}, U. (2004).
\newblock {Reference System, Conventions and Notations for Gaia}.
\newblock Research Note GAIA-ARI-BAS-003, GAIA livelink.

\bibitem[{Bini} et~al., 2003]{2003CQGra..20.4695B}
{Bini}, D., {Crosta}, M.~T., and {de Felice}, F. (2003).
\newblock {Orbiting frames and satellite attitudes in relativistic astrometry}.
\newblock {\em Class.\ Quantum Grav.}, 20:4695--4706.

\bibitem[{Bini} and {de Felice}, 2003]{2003CQGra..20.2251B}
{Bini}, D. and {de Felice}, F. (2003).
\newblock {Ray tracing in relativistic astrometry: the boundary value problem}.
\newblock {\em Class.\ Quantum Grav.}, 20:2251--2259.

\bibitem[de~Felice et~al., 2001]{2001AAp...373..336D}
de~Felice, F., {Bucciarelli}, B., {Lattanzi}, M.~G., and {Vecchiato}, A.
  (2001).
\newblock {General relativistic satellite astrometry. II.\ Modeling parallax
  and proper motion}.
\newblock {\em Astron.\ Astrophys.}, 373:336--344.

\bibitem[de~Felice and {Clarke}, 1990]{1990recm.book.....D}
de~Felice, F. and {Clarke}, C.~J.~S. (1990).
\newblock {\em {Relativity on curved manifolds}}.
\newblock {Cambridge University Press}.

\bibitem[{de Felice} et~al., 2004]{2004ApJ...607..580D}
{de Felice}, F., {Crosta}, M.~T., {Vecchiato}, A., {Lattanzi}, M.~G., and
  {Bucciarelli}, B. (2004).
\newblock {A General Relativistic Model of Light Propagation in the
  Gravitational Field of the Solar System: The Static Case}.
\newblock {\em Astrophys.\ J.}, 607:580--595.

\bibitem[de~Felice et~al., 1998]{1998AAp...332.1133D}
de~Felice, F., {Lattanzi}, M.~G., {Vecchiato}, A., and {Bernacca}, P.~L.
  (1998).
\newblock {General relativistic satellite astrometry. I. A non-perturbative
  approach to data reduction}.
\newblock {\em Astron.\ Astrophys.}, 332:1133--1141.

\bibitem[IAU, 2000]{2000IAU-res....B1.3}
IAU (2000).
\newblock {Definition of Barycentric Celestial Reference System and Geocentric
  Celestial Reference System}.
\newblock IAU Resolution B1.3 adopted at the 24th General Assembly, Manchester,
  August 2000.

\bibitem[{Jantzen} et~al., 1992]{1992AnnPhys.215..1J}
{Jantzen}, R.~T., {Carini}, P., and {Bini}, D. (1992).
\newblock {The many faces of gravitoelectromagnetism}.
\newblock {\em Ann. Phys.}, 215:1--50.

\bibitem[{Klioner}, 2003]{2003AJ....125.1580K}
{Klioner}, S.~A. (2003).
\newblock {A Practical Relativistic Model for Microarcsecond Astrometry in
  Space}.
\newblock {\em Astron.\ J.}, 125:1580--1597.

\bibitem[{Klioner} and {Peip}, 2003]{2003A&A...410.1063K}
{Klioner}, S.~A. and {Peip}, M. (2003).
\newblock {Numerical simulations of the light propagation in the gravitational
  field of moving bodies}.
\newblock {\em Astron.\ Astrophys.}, 410:1063--1074.

\bibitem[{Kopeikin} and {Mashhoon}, 2002]{2002PhRvD..65f4025K}
{Kopeikin}, S.~M. and {Mashhoon}, B. (2002).
\newblock {Gravitomagnetic effects in the propagation of electromagnetic waves
  in variable gravitational fields of arbitrary-moving and spinning bodies}.
\newblock {\em Phys.\ Rev.\ D}, 65:64025.

\bibitem[{Kopeikin} and {Sch{\" a}fer}, 1999]{1999PhRvD..60f124002K}
{Kopeikin}, S.~M. and {Sch{\" a}fer}, G. (1999).
\newblock {Lorentz Covariant Theory of Light Propagation in Gravitational
  Fields of Arbitrary Moving Bodies}.
\newblock {\em Phys.\ Rev.\ D}, 60:124002.

\bibitem[{Le Poncin-Lafitte} et~al., 2004]{2004CQGra..21.4463L}
{Le Poncin-Lafitte}, C., {Linet}, B., and {Teyssandier}, P. (2004).
\newblock {World function and time transfer: general post-Minkowskian
  expansions}.
\newblock {\em Class.\ Quantum Grav.}, 21:4463--4483.

\bibitem[{Misner} et~al., 1973]{1973grav.book.....M}
{Misner}, C.~W., {Thorne}, K.~S., and {Wheeler}, J.~A. (1973).
\newblock {\em Gravitation}.
\newblock San Francisco: W.H.~Freeman and Co.

\bibitem[{Turon} et~al., 2005]{2005tdug.conf.....T}
{Turon}, C., {O'Flaherty}, K.~S., and {Perryman}, M.~A.~C., editors (2005).
\newblock {\em {The Three-Dimensional Universe with Gaia}}.

\bibitem[{Weinberg}, 1972]{1972gcpa.book.....W}
{Weinberg}, S. (1972).
\newblock {\em {Gravitation and Cosmology: Principles and Applications of the
  General Theory of Relativity}}.
\newblock Gravitation and Cosmology: Principles and Applications of the General
  Theory of Relativity, by Steven Weinberg, pp.~688.~ISBN
  0-471-92567-5.~Wiley-VCH , July 1972.

\end{thebibliography}
\end{document}